\documentclass[aps,prx,preprint,groupedaddress,amsmath,amssymb,longbibliography]{revtex4-1}

\usepackage{graphicx}  
\usepackage{dcolumn}   
\usepackage{bm}        
\usepackage{verbatim}   
\usepackage{subfigure}
\usepackage{amsmath}
\usepackage{natbib}
\usepackage{xcolor}
\usepackage{lipsum} 
\usepackage{hyperref}

\newcommand{\CV}{\operatorname{\mathrm{CV}}}
\newcommand{\erf}{\operatorname{\mathrm{erf}}}

\usepackage[draft]{todonotes}   
\usepackage{color}
\usepackage{graphicx}  
\usepackage{dcolumn}   
\usepackage{bm}        
\usepackage{verbatim}   
\usepackage{subfigure}
\usepackage{amsmath}
\usepackage{amsfonts}
\usepackage{amssymb}
\usepackage{graphics}
\usepackage{pgf,pgffor}
\usepackage{hyperref}

\begin{document}

\title{Joint Statistics of Strongly Correlated Neurons via Dimensional Reduction}
\author{Ta\c{s}k{\i}n Deniz \& Stefan Rotter}
\affiliation{
	Bernstein Center Freiburg \& Faculty of Biology,
	University of Freiburg,
	Hansastra{\ss}e 9a, 79104 Freiburg, Germany
}
\date{\today}

\begin{abstract}
	The relative timing of action potentials in neurons recorded from local cortical networks often shows a non-trivial dependence, which is then quantified by cross-correlation functions. Theoretical models emphasize that such spike train correlations are an inevitable consequence of two neurons being part of the same network and sharing some synaptic input. For non-linear neuron models, however, explicit correlation functions are difficult to compute analytically, and perturbative methods work only for weak shared input. In order to treat strong correlations, we suggest here an alternative non-perturbative method. Specifically, we study the case of two leaky integrate-and-fire neurons with strong shared input. Correlation functions derived from simulated spike trains fit our theoretical predictions very accurately. Using our method, we computed the non-linear correlation transfer as well as correlation functions that are asymmetric due to inhomogeneous intrinsic parameters or unequal input.   
\end{abstract}

\maketitle

\section{Introduction}

Electric activity generated by different neurons in the brain is often strongly correlated \cite{Rosenbaum2014, Lampl1999,Okun2008,Poulet2008}. These correlations arise as a result of shared input, or input components that are themselves correlated. Correlated activity can be a consequence of shared background fluctuations \cite{Arieli1996}, but strong correlations might also indicate synchronous action potentials at the input indicating temporal coding. However, a clear-cut dichotomy between decorrelated and synchronized dynamics does not exist \cite{Staude2008} \cite{Kumar2010, Doiron2016a}. Rather, one should consider these two extremes as two faces of the same coin. Recent high-precision measurements reported very low average correlations suggesting a mechanism of active decorrelation in cortical networks \cite{Ecker2010,Renart2010,Pernice2011,Helias2014}. At the same time it was observed by intracellular measurements that nearby neurons receive very similar input \cite{Lampl1999,Okun2008,Poulet2008}. 

Several studies of pair correlations in neural networks relate structure and dynamics assuming a fluctuating dynamics about a fixed point that is characterized by asynchronous (A) population activity and irregular (I) spike trains \cite{Brunel2000, Pernice2011, Pernice2012, Trousdale2012}. They employ essentially linear perturbation theory \cite{Brunel1999,Lindner2001} to compute correlation functions. Nevertheless, some of these works push the limits of existing methods. First of all, a qualitatively different AI state was observed in simulations of spiking neural networks with stronger couplings \cite{Ostojic2014}. Secondly, a partial extension of the theory to the strongly correlated regime was based on numerically determined spike response functions \cite{Pernice2012}. Thirdly, pair correlation studies were generalized to higher-order correlations in recurrent networks by accounting for certain network connectivity motifs \cite{Jovanovic2015,Jovanovic2016}. These studies exploit and extend existing methods, but they also demonstrate the need for a new approach. 

The main assumption of perturbation theory is that the common drive of the two neurons is weak. Yet, this criterion depends on the background state and strength of interactions in a given network. We showed previously that low background rates, for example, may lead to a breakdown of perturbation theory even for low correlation coefficients $c$ \cite{Deniz2016}. This makes non-perturbative methods indispensable for modeling and analysis of correlations as low spike rates are typical in experiments \cite{Shadlen1998}. All in all, a proper treatment of strong correlations must take the non-linear correlation transfer into account, which appears to play an important role in sensory processing \cite{Lyamzin2015}. However, a unified and transparent framework to calculate correlations of all strengths for neuron models of the integrate-and-fire type does not exist. 

The pitfall of previously suggested non-perturbative methods are their immense computational costs due to the high dimensionality of the discretized problem \cite{Rosenbaum2012}. This makes computations practically impossible for a large range of parameters. For instance, the numerical effort of computing the pair correlation problem scales as $N^4$, where $N$ is the number of grid points used to approximate single neuron membrane potentials. Although, limiting the grid size is possible \cite{Helias2010}, a too coarse voltage grid fails to properly reflect the statistics of leaky-integrate-and-fire neurons with Poisson input. The precision issue gets even more severe for correlations of higher order, which are needed to parametrize the joint statistics of multiple neurons. With our methods, in contrast, we observed that joint membrane potential distributions of even strongly correlated neurons can be reduced to a small set of principal vectors via singular value decomposition (SVD). This suggests that strong correlations can be computed with high precision resorting to subspaces of relatively low dimension. In this work, specifically, we devise a SVD based method that allows to compute spike correlation functions of two leaky integrate-and-fire neurons with high accuracy.

Similar problems were studied analytically for arbitrary input correlations of the stochastic dynamics of neural oscillators \cite{Abouzeid2011}, and for level-crossings of correlated Gaussian processes \cite{Tchumatchenko2010a}. Related numerical work considered strong input correlations for integrate-and-fire neurons receiving white noise input \cite{Vilela2009} or shot noise input with nontrivial temporal correlations \cite{Schwalger2015, Voronenko2015}. The problem of analytically calculating the stationary distributions conditional on a spike from the exit current at the threshold is also discussed in the case of colored noise \cite{Schwalger2015}. A method to deal with very strong input correlations $(c\approx 1)$ in a specific input model (correlated Poisson processes) was suggested by \cite{Schultze-Kraft2013}. Our study further suggests a novel technique, extending \cite{Helias2010}, to solve 2D jump equations for leaky integrate-and-fire neurons by mapping it to a Markov chain. This method provides an accurate estimate of the steady state joint distribution of membrane potentials.
   
We test our method for large input correlations $c$ and demonstrate its power for different types of correlation asymmetries by comparing our semi-analytical approach to correlations extracted from simulated neuronal spike trains. We look at the full range of input correlations and provide an example of a non-linear correlation transfer function. Our method can be extended to more general integrate-and-fire models, higher-order input correlations and third order output correlations. However, we have to defer a detailed analysis of such cases to future work. 

\begin{figure}[h!]
	\centering
	\includegraphics[width=10cm]{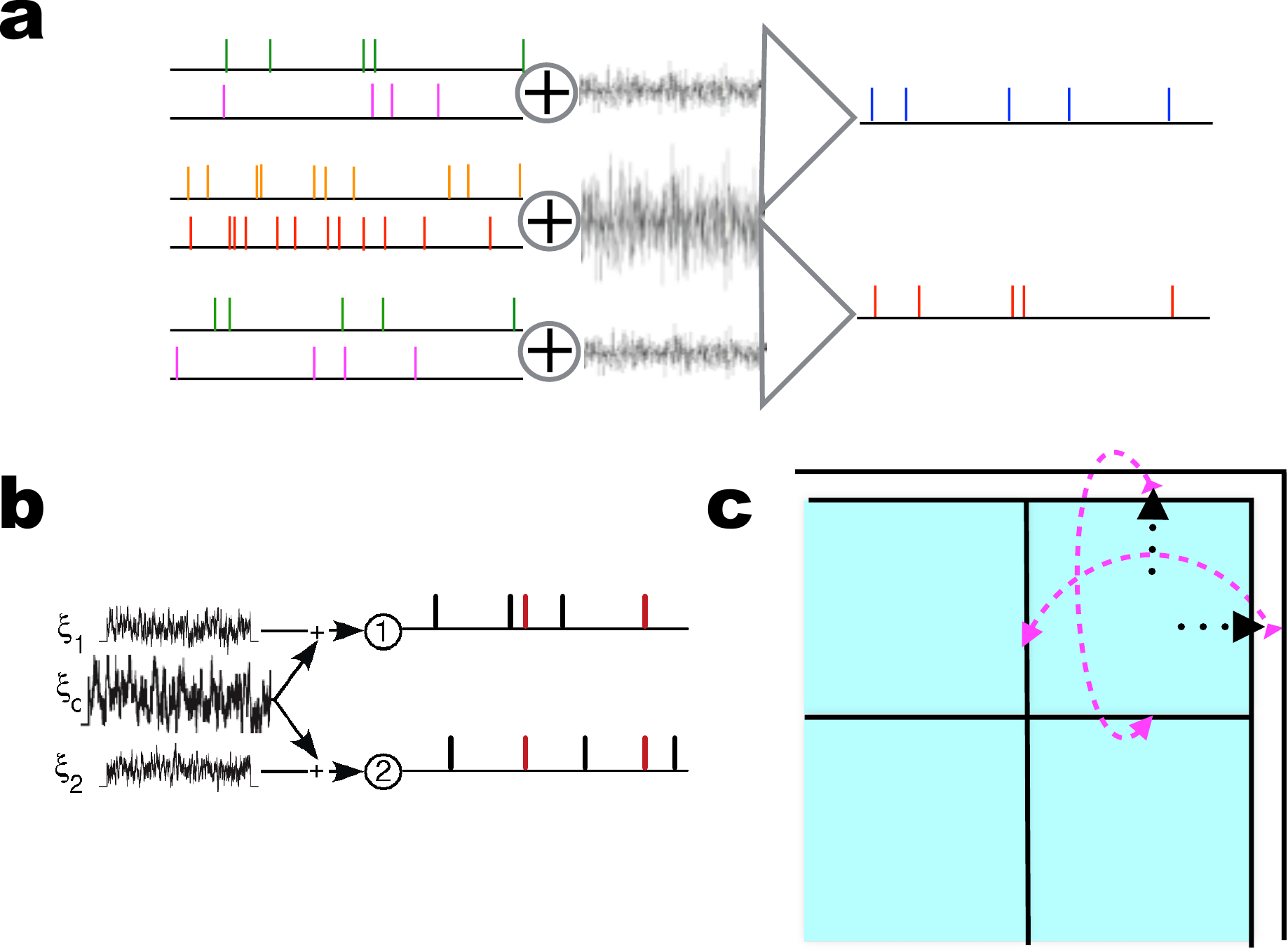}
	\caption{(a)~Two LIF neurons receiving private and shared shared inputs, both represented by excitatory and inhibitory spike trains. (b)~Two neurons with shared white noise input, with first and second moments matched to (a). (c)~Schematic describing the threshold crossing and reset mechanism that is part of the membrane potential dynamics of LIF neurons.}
	\label{fig_1}
\end{figure}

\section{Models \& Methods}

\subsection{Two neurons with shared Poisson input}\label{sec:two-neurons-with-shared-poisson--input}

The leaky inteagrate-and-fire neuron model with postsynaptic potentials of finite amplitudes was studied previously in \cite{Helias2010}. The stochastic equation that describes the membrane potential dynamics of one particular neuron is given as 
\begin{equation}
\tau_m\dot{V}(t)=-V(t)+h_{ex}\tau\sum_i S^{ex}_i(t) +h_{in} \tau_m \sum_j S^{in}_j(t). 
\end{equation}
where $h_{ex}$ and $h_{in}$ represent the amplitudes of individual EPSPs and IPSPs, respectively, and $S^{ex}_i(t)$ and $S^{in}_j(t)$ are the spike trains of excitatory and inhibitory presynaptic neurons, with each of their spikes represented by a Dirac delta function
\begin{equation}
S^{ex}_i(t)=\sum_{n}\delta(t-t^{i}_n).
\end{equation}
An analogous definition holds for inhibitory neurons. In both cases, if the membrane potential reaches the firing threshold, $V_{th}$, a spike is elicited and the voltage is reset to its resting value at $0$.

In order to study correlations between the spike trains of two neurons we look at two coupled stochastic equations, describing the membrane potentials of two neurons that share a certain fraction of their excitatory and inhibitory input spikes
\begin{align}
\dot{V_1}(t)&=-\frac{V_1(t)}{\tau_{m,1}}+h_{ex}(\sum_k S^{ex}_{1,k}(t)+\sum_m S^{ex}_{m}(t) )+
h_{in}(\sum_l S^{in}_{1,l}(t) +
\sum_n S^{in}_{n}(t) )\\
\dot{V_2(t)}&=-\frac{V_2(t)}{\tau_{m,2}}+h_{ex}(\sum_k S^{ex}_{2,k}(t)+\sum_m S^{ex}_{m}(t) )+
h_{in}(\sum_l S^{in}_{2,l}(t) +
\sum_n S^{in}_{n}(t) )
\label{eq_model}
\end{align}
where shared excitatory and inhibitory input spike trains are denoted as $\sum_m S^{ex}_{s,m}(t)$ and $\sum_n S^{in}_{s,n}(t) $, respectively. 
Comparing the parameters of the jump model to shared white noise input we implicitly specified the firing rates of 6 independent Poisson processes 
(Fig.~\ref{fig_1})
\begin{align}
\dot{V_1}(t)&=-\frac{V_1(t)}{\tau_{m,1}}+\mu_1+\frac{\sigma_1}{\sqrt{\tau_{m,1}}} \xi_1(t)+ \frac{\sigma_{c,1}}{\sqrt{\tau_{m,1}}} \xi_c(t)\\
\dot{V_2(t)}&=-\frac{V_2(t)}{\tau_{m,2}}+\mu_2+\frac{\sigma_2}{\sqrt{\tau_{m,2}}}  \xi_2(t)+ \frac{\sigma_{c,2}}{\sqrt{\tau_{m,2}}}   \xi_c(t).
\label{eq_dif_model}
\end{align}
For notational convenience, shared input rates $r_{ex,c}$ and $r_{inh,c}$ are computed from a shared Wiener process $\xi_c$ with zero mean. Setting $h_{ex}=h$ and $h_{in}=gh$, rates for each independent excitatory and inhibitory process are given as 
\begin{align}
r_{ex}&=\frac{1}{1+g}\Big{(}\frac{\sigma^2}{\tau h^2}+\frac{g\mu}{h\tau}\Big{)}\\
r_{in}&=\frac{1}{(g+g^2)}\Big{(}\frac{\sigma^2}{\tau h^2}-\frac{\mu}{h\tau}\Big{)}.
\end{align}
Here we note that some combinations of $(\mu,\sigma)$ do not correspond to any combination of positive Poisson rates, as they must satisfy
\begin{eqnarray}
\sigma^2\geq h\mu
\end{eqnarray}
guarantee $r_{in}\geq 0$ for the inhibitory rate.

\begin{figure}[h!]
	\begin{center}
		\includegraphics[width=8cm]{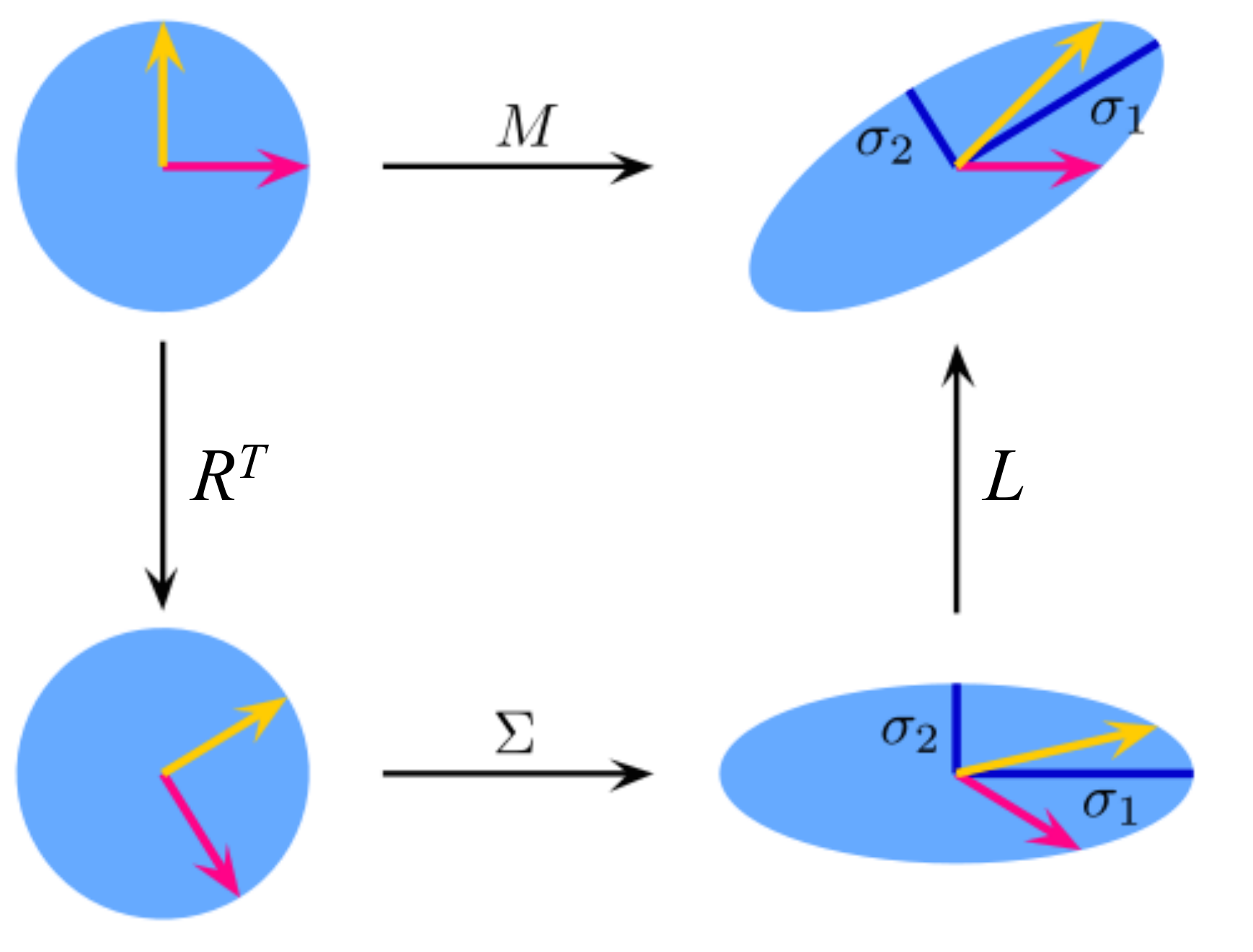}
	\end{center}
	\caption{Schematic illustration of Singular Value Decomposition in 2 dimensions, $M=L \Sigma R^T$, for positive definite matrices ($\det(M)>0$). $L$ and $R$ are orthogonal matrices, and $D$ is a real diagonal matrix. The non-negative diagonal elements $\sigma_1$ and $\sigma_2$ of the matrix $\Sigma$ are the so-called singular values of the matrix $M$. }
	\label{fig2}
\end{figure}

\subsection{Discretized Markov operators for the LIF model}

In this section we summarize the discrete approximation to the dynamics of a LIF neuron, as developed in \cite{Helias2010}. The coarse graining of the membrane potential is given by the map 
\begin{equation}
\mathbb{R} \rightarrow \mathbb{Z}, \quad  V \mapsto \lfloor \frac{V}{\Delta V }\rfloor .
\end{equation}
The probability density function $P(v)$ of the membrane potential then becomes a vector $p = (p_i)$ satisfying
\begin{equation}
p_i = \int^{v_{i+1}}_{v_i} P(v)\,dv,
\end{equation}
with $v_i=i \, \Delta V $. As we impose a cut-off lower boundary $V_-$ of the voltage scale, the dimension of the discrete state space  is given as $N=\frac{V_{th}-V_-}{\Delta V}$.

The temporal evolution of the membrane potential distribution is now described in terms of a Markov process. The Markov propagator can be expressed as a juxtaposition of three operators: a decay operator $\mathcal{D}$ describing leaky integration, a jump operator $\mathcal{J}$ accounting for the action of synaptic inputs, and a threshold-and-reset operator $\mathcal{T}$ that implements spike generation upon threshold crossing.
The discrete decay operator $\mathcal{D}$ is derived from its continuous counterpart $D$ as follows
\begin{multline*}
\Delta V \mathcal{D}\vec{p}= \int^{v_{i+1}}_{v_i} DP(x)\,dx = \int^{v_{i+1}}_{v_i} \underbrace{e^{\Delta t/\tau_m}}_qP(e^{\Delta t/\tau_m}x)\,dx\\
= \sum_{\lfloor qi\rfloor}^{\lceil q(i+1) \rceil-1} p_j.\Delta V+
\Big{(} \Delta V \lceil qi \rceil-qv_i \Big{)} p_{\lfloor q_i\rfloor}+
\Big{(}qv_{i+1}-\Delta V \lfloor q(i+1)\rfloor \Big{)} p_{ \lfloor q(i+1)\rfloor}.
\end{multline*}
This definition leads to the following decay matrix
\begin{equation}
\mathcal{D}_{ij}= \sum_{\lfloor qi\rfloor}^{k=
	\lceil q(i+1) \rceil-1}
\delta_{j,k}+ \Big{(} \lceil qi \rceil-qv_i \Big{)} \delta_{j,{\lfloor q_i\rfloor}}+
\Big{(}qv_{i+1}-\lfloor q(i+1)\rfloor \Big{)} \delta_{j,{ \lfloor q(i+1)\rfloor}}.
\end{equation}
The jump distribution, which underlies the jump operator $\mathcal{J}$, can be derived from the count distribution of excitatory and inhibitory synaptic events in a small time interval $\Delta t$. Assuming that they arrive with Poisson statistics (maximum entropy) at a rate of $r_{ex}$ and $r_{in}$, respectively, we obtain
\begin{equation}
\label{eq_P_q_r}
P(m,n)=\frac{a^m \: b^n}{m! \: n!} e^{-(a+b)}
\end{equation}
where $a= r_{ex}\Delta t$ and $b=  r_{in}\Delta t$ are the corresponding expected event counts. Dummy indices $m$ and $n$ correspond to excitatory and inhibitory counts. The jump distribution of the membrane potential $\gamma=(m-gn)h$ is given by
\begin{equation}
\label{eq_jump_dist}
P(\gamma)=\sum_{m,n=0}^{\infty}P(m,n)\delta_{\gamma,(m-gn)h}.
\end{equation}
The jump operator $\mathcal{J}$ is then derived as
\begin{equation*}
(\mathcal{J}p)_{i}=\sum_{\gamma}  P(\gamma) p_{i-\frac{\gamma}{\Delta V}} 
\end{equation*}
with a jump matrix given by
\begin{equation}
\mathcal{J}_{ij}=\sum_{\gamma}  P(\gamma) \delta_{j,i-\frac{\gamma}{\Delta V}} .
\end{equation}
The threshold-and-reset operator $\mathcal{T}$ takes all the states above threshold and inserts them at the reset potential. This is simply given as 
\begin{equation}
\mathcal{T}_{ij}=I_{j>\frac{V_{th}}{\Delta V }} \delta _{i,\frac{V_{r}}{\Delta V }}+I_{j\leq \frac{V_{th}}{\Delta V }} \delta _{i,j}
\label{eq_threshold_matrix}
\end{equation}
Finally, the time evolution matrix of the corresponding Markov chain is given as a product of the three operators described above
\begin{equation}
\mathcal{M}=\mathcal{T}\mathcal{J}\mathcal{D}.
\label{eq_M1d}
\end{equation}
The discrete stationary distribution is 
\begin{equation}
\mathcal{M}P_0=P_0,
\end{equation}
and the corresponding stationary rate is given as 
\begin{align}
\label{eq_rate_markov}
r_{\mathrm{Markov}}&=(\tau_{\mathrm{ref}}+[\frac{1}{h} \sum_{i \geq i_{th}} (\mathcal{J}\mathcal{D}P_0)_i]^{-1})^{-1}.
\end{align}

\subsection{Spike train correlations}

The cross-covariance function of two stationary spike trains $S_a(t) = \sum_l \delta(t-t^a_l)$ ($a = 1,2$) is defined as
\begin{equation}
	C_{12} (\tau)= \langle S_1(t+\tau) S_2(t) \rangle 
	-\langle S_1(t+\tau)\rangle \langle S_2(t) \rangle 
\end{equation}
where $\langle S_a(t)\rangle = r_a$, with $\langle .\rangle$ indicating the ensemble average.
For the model we studied here stationary rates were computed using Eq.~\ref{eq_corr_rate}.
The cross-covariance function can be expressed in terms of the conditional firing rate $r_{1|2}(\tau)$, two stationary rates $r_1$ and $r_2$, and the amplitude $r_0$ of a $\delta$-function
\begin{equation}
	\label{eq_corr_rate}
	C_{12} (\tau)= r_0\delta(\tau)+r_2 (r_{1|2}(\tau)-r_1).
\end{equation}
Given the spiking neuron model considered here, Eq.~\ref{eq_corr_rate} can be derived from the stationary joint membrane potential distribution $P_{0}(V_1,V_2)$.
Conditional on a spike at time $t_0$ in the first neuron, one has to find the instantaneous distribution of the membrane potential of the second neuron 
\begin{equation}
	P_{i|j}(V_i)=P(V_i \: | \: \text{neuron $j$ spikes at $t_0$}).
\end{equation}
The probability of observing a consecutive spike is given by the flux  $P_{\mathrm{flux}}(V_1,\theta_2)$ with the normalization in Eq.~\ref{eq_IVP}.  The conditional flux is computed for the Markov approximation as in Eq.~\ref{eq_flux}.
In general we simply compute the exit rate at the threshold distributed over $V_1$ by solving the following initial value problem 
\begin{align}
	\label{eq_IVP}
	P(0,V_1)&=P_{1|2}(V_1)=P_{\mathrm{flux}}(V_1,\theta_2)\Big{(}\int_{\infty}^{\theta_1} P_{\mathrm{flux}}(x,\theta_2) \,dx\Big{)}^{-1}\\
	\Delta_t P(t)&=\mathcal{M}_1 P(t)
\end{align}
where $\mathcal{M}_1$ is discrete time evolution matrix of the neuron model given by Eq.~\ref{eq_M1d}. $\mathcal{M}_1$ leads to a (forward) time evolution in steps of $\Delta t $. The instantaneous conditional rate $r_{1|2}(t)$  in Eq.~\ref{eq_corr_rate} is computed with a formula similar to Eq.~\ref{eq_rate_markov} 
\begin{align}
\label{eq_rate_markov_t}
r_{1|2}(t)=r_{\mathrm{Markov}}(t)&=(\tau_{\mathrm{ref}}+[\frac{1}{h} \sum_{i \geq i_{th}} (\mathcal{J}\mathcal{D}P(t))_i]^{-1})^{-1}.
\end{align}
Finally, the covariance function $C_{21} (\tau)$ in Eq.~\ref{eq_corr_rate}  is derived by using $r_{1|2}(t)$ and two stationary rates $r_1$ and $r_2$ and $r_0$.  Note that the method described here can be generalized to higher order correlations as well. 


\subsection{Correlated jump distribution in 2D}

We now use a 2-dimensional state space  describing the joint membrane potential evolution of two neurons. Correlated and uncorrelated Poisson jumps push the 2D membrane potential vector $(V_1,V_2)$ into three independent directions, $(1,0)$, $(0,1)$ and $(1,1)$, allowing jumps in the positive (excitatory) and negative (inhibitory) direction, respectively. Hence, the jump distribution from an initial position $(U_1,U_2)$ to a new position $(V_1,V_2)$ in state space  driven by $6$ independent Poisson processes is obtained from a 2D convolution 
\begin{equation}
	P(V_1,V_2|U_1,U_2)=\int P_{1}(V_1-U_1-Z)P_{2}(V_2-U_2-Z)P_{c}(Z)\,dZ.
	\label{P_jump_2d}
\end{equation}
Inserting the expressions for the jump distributions in each direction, collecting all terms and using Eq.~\ref{eq_P_q_r}, we obtain
\begin{multline*}
	P(V_1,V_2)=\int \sum_{i,j=0}^{\infty}\sum_{k,l=0}^{\infty}\sum_{m,n=0}^{\infty} P_{1}(i,j)\\
	\times P_{2}(k,l)  P_{c}(m,n)\delta_{x-Z,(i-gj)h}\delta_{y-Z,(k-gl)h}\delta_{Z,(m-gn)h} \,dZ.
\end{multline*}
This expression is also valid for more general input statistic models that rely on a decomposition into statistically independent parts \cite{Kuhn2003,Schultze-Kraft2013}. Here we consider the shared Poisson input model as described by Eq.~\ref{eq_model}. Integrating the expression with respect to $Z$ and inserting the mean event counts $a_i= r_{ex,i}\Delta t$ and $b_i=  r_{in,i}\Delta t$, the resulting expression is given in a compact form
\begin{multline*}
	P(V_1,V_2)=\sum_{i,j=0}^{\infty}\sum_{k,l=0}^{\infty}\sum_{m,n=0}^{\infty} e^{-(a_1+b_1+a_2+b_2+a_s+b_s)} \\
	\times \frac{a_1^i \: b_1^j}{i! \: j!} \frac{a_2^k \: b_2^l}{k! \: l!} \frac{a_s^m \: b_s^n}{m! \: n!} 
	\: \delta_{V_1,(i-m-g(j-n))h} \:\delta_{V_2,(k-m-g(l-n))h}.
	\label{P_jump_2d}
\end{multline*}
We can simplify this expression by choosing a regular grid according to $h_e= n  \Delta V$ and $h_i= m \Delta V $, for integers $m$ and $n$.  
In practice, however, the resulting sum will be truncated for a given tolerance. We will use the matrix form of the discretized operators in the following subsection, which is equivalent to the above expression. 

\subsection{Linear operators for correlated dynamics in 2D}

Here we discuss the action of operators on state vectors of the discretized $(V_1,V_2)$ space, assuming $N$ bins in each dimension. We write the stationary voltage distribution in the form
\begin{equation}
\label{eq_p_0}
	P_0(X,Y) \equiv \sum_{i,j=1}^N \Omega_{ij}  \: X_i\otimes Y_j.
\end{equation}
where $X$ and $Y$ are two suitable orthogonal bases of $\mathbb{R}^N$. We will define a specific choice for X and Y later in Section~\ref{SVD}. 
As there is no crosstalk between the two neurons except by shared input leading to a correlated jump distribution, threshold and decay operators (tensors) are separable
\begin{align}
	\mathcal{D}_{2D}&=\mathcal{D}_1\otimes \mathcal{D}_2 \\
	\mathcal{T}_{2D}&=\mathcal{T}_1\otimes \mathcal{T}_2.
\end{align}
Separability would also apply to the jump distribution for an input correlation coefficient of $c=0$, corresponding to two neurons without shared input
\begin{equation}
	\mathcal{J}_{2D}=\mathcal{J}_1\otimes \mathcal{J}_2.
\end{equation}
However, in the case of non-zero correlation, this relation holds only for a single path among the many connecting two points in state space . Therefore, every path must be taken into account by considering the contribution of each operator to a movement in the oblique $(1,1)$ direction. 

Once we have computed the correct Markov matrix for 2D via
\begin{equation}
	\mathcal{M}_{2D}=\mathcal{T}_{2D} \: \mathcal{J}_{2D} \: \mathcal{D}_{2D}
	\label{eq_M2d}
\end{equation}
one can also find the stationary joint membrane potential distribution as the Perron-Frobenius eigenvector $P_0$ of the propagator matrix $\mathcal{M}_{2D}$
\begin{equation}
	\mathcal{M}_{2D} P_0=P_0.
	\label{eq_full_markov}
\end{equation}
Regarding the correlated jump distribution there are two ways of constructing $\mathcal{J}$ operators. One possibility is described in Eq.~\ref{P_jump_2d}. Alternatively, we construct linear Markov jump operators in 2D exploiting the commutativity of infinitesimal operators
\begin{equation*}
	\mathcal{J}_{2D}=e^{(J^p_{1} \otimes I +I\otimes  J^p_{2}+J^c_{1} \otimes J^c_{2})}.
\end{equation*}
Here $I$ is the identity matrix. Using the properties of the operator product $\otimes$ and commutativity of the individual factors, we can simplify this expression
\begin{equation*}
	\mathcal{J}_{2D}=(e^{J^p_{1}} \otimes I)(I \otimes e^{J^p_{2}}) (e^{J^c_{1} \otimes J^c_{2}}).
\end{equation*}
In order to expand the third term we define $U$ and $D$ operators as up and down transition matrices, where
$U$ corresponds to a $1$-step up transition, and $D$ corresponds to a $1$-step down transition. This leads to
\begin{equation}
	U^{k}_{ij}=\delta_{i,j-k} 
	\quad\text{and}\quad
	D^{l}_{ij}=\delta_{i,j+l} .
\end{equation}
Hence,  discrete approximations to infinitesimal generators of private components are given as 
\begin{equation}
J_i=a_iU^{k}+b_iD^{l}. 
\end{equation}
where matrix powers $k$ and $l$ are defined on  $h_e= k  \Delta V$ and $h_i= l \Delta V $ for simplicity.(This restrictive assumption can be generalized easily by computing the continuous jump distribution in Eq.~\ref{eq_jump_dist} and then discretizing it, which leads to the same result.) On the other hand, we need to be careful with correlated spikes, which trigger jumps into the oblique direction $(1,1)$ with probability 
\begin{equation}
c_{mn}=e^{-(a_c+b_c)}\frac{1}{m!}\frac{1}{n!}a_c^mb_c^{n}
\end{equation} 
for $m$ excitatory and $n$ inhibitory jumps. Expanding yields
\begin{equation}
	\label{eq_J_2D_1}
	\mathcal{J}_{2D}=\sum_{m=0} \sum_{n=0}c_{mn} [e^{J_1}U^{km} D^{ln}\otimes e^{J_2}U^{km} D^{ln}].
\end{equation} 
As we noted before, the above construction is quite general and can be applied easily for general amplitude distributions.  We only need to consider a discrete amplitude distribution $c_{mn}$ in a given time bin of size $\Delta t$, as described above, see also \cite{ Kuhn2003, Richardson2010}. 

A final remark on the method described in this section concerns the commutativity of matrices. This property leads to a numerically more economic expression
\begin{equation}
	\label{eq_J_2D}
	\mathcal{J}_{2D}=\sum_{j \: \in \: \mathbb{J}} c(j) [e^{J_1}O^{j}\otimes e^{J_2}O^{j}]
\end{equation} 
where the set of integers $\mathbb{J}$ is defined as list of all jump numbers $j=mk-nl$. The coefficient
\begin{equation}
c(j)= \sum_{m,n} e^{-(a_c+b_c)} c_{mn}\delta_{j,mk-nl}
\end{equation}
is the probability of $j$ jumps. The jump generator $O$ is then defined in terms of matrix powers
\begin{equation}
O^j=\begin{cases}
U^{j}, & j>0\\
I ,& j=0\\
D^{j}& j<0\\
\end{cases}.
\end{equation}

\subsection{Operator decomposition and SVD basis reduction}
\label{SVD}

The expansion method described above is straightforward, but rather cumbersome to implement. We will now introduce a basis for the expansion of correlated jump operators suitable to reduce the cost of the computations involved, and helpful to increase the accuracy of a truncation. With our method, as compared to others, we have to solve linear equations of lower dimensionality in order to get a better approximation for the correlation function. Some further arguments for selecting Singular Value Decomposition are discussed in the results section. SVD of a matrix is given as
\begin{equation}
	\mathcal{M}=\mathcal{L}\Sigma \mathcal{R}^T
\end{equation}
where the diagonal entries of the diagonal matrix $\Sigma$ are the square roots of the non-zero eigenvalues of $\mathcal{M}\mathcal{M}^T$ and $\mathcal{M}^T \mathcal{M}$. Both matrices $\mathcal{R}$ and $\mathcal{L}$ are orthogonal with columns consisting of the eigenvectors of $\mathcal{M}\mathcal{M}^T$ and $\mathcal{M}^T \mathcal{M}$, respectively. We show a 2D example SVD of Markov matrix in Fig.~\ref{fig2}.  For $\mathcal{M}$ replaced by a single-neuron time evolution matrix $\mathcal{M}_1$ ($\mathcal{M}_2$), we define the matrix $X$ ($Y$) by the selected orthogonal subspace of dimension $K$ ($L$) of $\mathcal{R}_1$ ($\mathcal{R}_2$), according to the largest $K$ ($L$) singular values, resepectively. In order to project $\mathcal{J}_1$ ($\mathcal{J}_2$)and $\mathcal{J}_2$ ($\mathcal{T}_2$) to this subspace, we also extend the orthogonal basis $X$ and $Y$ to supra-threshold transitions
\begin{equation}
	\tilde{X}=
	\begin{pmatrix}
		X & 0\\
		0 & I_{m}
	\end{pmatrix},
\qquad
	\tilde{Y}=
	\begin{pmatrix}
		Y & 0\\
		0 & I_{n}
	\end{pmatrix},
\end{equation}
where $X$ is an $M\times K$ and $Y$ is an $N\times L$ operator, respectively. $I_{m}$ and $I_{n}$ are identity matrices, where $m$ and $n$ are the maximal supra-threshold jump numbers induced by both private and shared inputs. The combined action of $\mathcal{J}$ and $\mathcal{T}$ for $c=0$ is then expressed as reduced operators 
\begin{equation}
\label{eq_reduced_operators}
\mathcal{T}_1 J_1 \rightarrow X^T \mathcal{T}_1\tilde{X} \tilde{X}^T \mathcal{J}_1 X,
\qquad
\mathcal{T}_2 J_2 \rightarrow Y^T \mathcal{T}_2 \tilde{Y} \tilde{Y}^T \mathcal{J}_2 Y,
\end{equation}
which map $M\times M$ ($N\times N$) to $K \times K$ ($L \times L$) matrices. Below we use the same dimensional reduction for correlated operators. In Eq.~\ref{eq_J_2D} the correlated jump operators are expressed as
\begin{equation*}
\mathcal{J}_{2D}=	\sum_{j \: \in \: \mathbb{J}}c(j) \: A_j \otimes B_j.
\end{equation*}
A dimensional reduction is then achieved by using Eq.~\ref{eq_reduced_operators} in
\begin{equation*}
\mathcal{M}_{2D}P_0=	\sum_{j \: \in \: \mathbb{J}} c(j) \: (\mathcal{T}_1 A_j \mathcal{D}_1\otimes \mathcal{T}_2 B_j\mathcal{D}_2 ) P_0.
\end{equation*}
In order to find $P_0$ in Eq.~\ref{eq_p_0} we need to solve Eq.~\ref{eq_full_markov} , which reads
\begin{equation*}
\mathcal{M}_{2D}P_0=P_0.
\end{equation*}
The projection operators $X^T\otimes Y^T$ satisfy $(X \otimes Y) (X^T \otimes Y^T)P_0=P_0$. Applying them to the left hand side of this equation Eq.~\ref{eq_full_markov}, we obtain 
\begin{align*}
(X^T \otimes Y^T)\mathcal{M}_{2D}&=(X^T \otimes Y^T)\mathcal{M}_{2D}(X \otimes Y) (X^T \otimes Y^T)P_0\\
&=\sum_{j \: \in \: \mathbb{J}} c(j) \:\Big{(}(X^{T}\mathcal{T}_1\tilde{X} \tilde{X}^T A_j X X^T\mathcal{D}_1 X) \otimes (Y^T \mathcal{T}_2 \tilde{Y} \tilde{Y}^T B_j Y \mathcal{D}_2 Y )\Big{)} \Omega\\
&\stackrel{RHS}{=}\Omega.
\end{align*}
$\Omega\equiv (X^T  \otimes Y^T)P_0$ can be expressed in a simpler form as
\begin{align*}
(X^T\otimes Y^T) P_0&=(X^T\otimes Y^T) (\sum_{ij} \Omega_{ij}X_i \otimes Y_j)\\
&=\sum_{ij}\Omega_{ij} X^TX_i \otimes Y^T Y_j)\\
&=\sum_{ij}\Omega_{ij} e_i \otimes e_j\equiv \Omega
\end{align*}
for monomials $(e_i)_k=\delta_{i,k}$. The reduced equation is then
\begin{equation}
\label{eq_reduced_eigen}
\mathcal{Q} \Omega=\Omega
\end{equation}
where $\mathcal{Q}$ is a tensor defined as
\begin{equation}
\label{eq_Q}
\mathcal{Q}=\sum_{j \: \in \: \mathbb{J}} c(j)(X^{T}\mathcal{T}_1\tilde{X} \tilde{X}^T A_j X X^T\mathcal{D}_1 X) \otimes (Y^T \mathcal{T}_2 \tilde{Y} \tilde{Y}^T B_j Y \mathcal{D}_2 Y^T Y).
\end{equation}
The dimensionally reduced problem in Eq.~\ref{eq_reduced_eigen} can then be solved in practice by reindexing tensor indices $(i,j,k,l) \mapsto (I,K) $ .


\begin{figure}
	\centering
	\includegraphics[width=12cm]{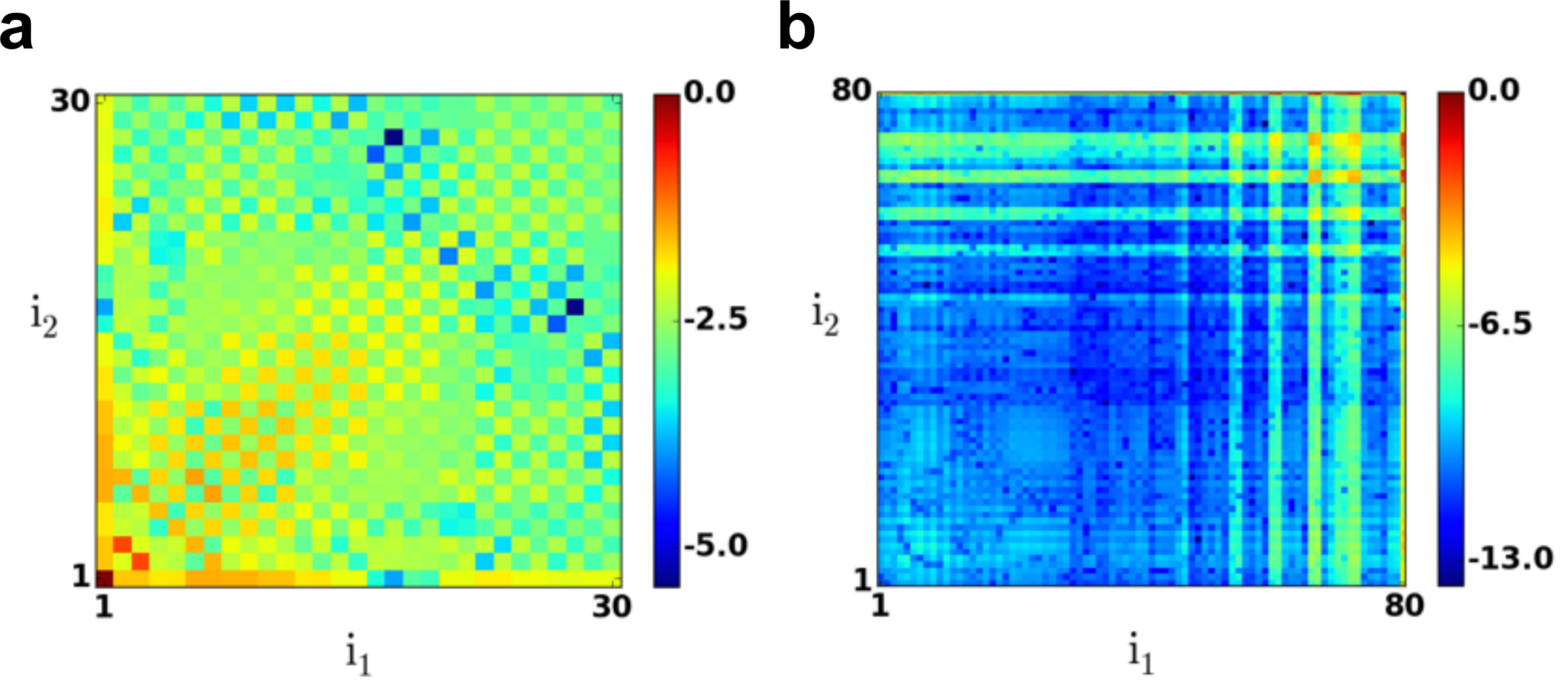}
	\caption{Direct projections suffer from the imposed lower boundary and diverging dual eigenvectors. Therefore, we cannot increase the precision of our method using direct projections, as demonstrated above (a)~ we show components $log_{10}(\Omega_{ij}/\max(S))$ with $N=30$ eigenvectors via dual space projections. (b)~Same as (a) with $N=80$ eigenvectors. We observed that the example in (b) fails to converge as its maximum value is $S_{80,80}$, because of numerical instabilities. }
	\label{fig3}
\end{figure}

\subsection{Conditional flux distribution}

Using the decomposed 2D stationary distribution obtained by reduction, one can compute the flux distribution with the help of matrix operators. We compute the flux distribution using the 2D decay and jump operators with thresholding imposed only at one of the boundaries. A scheme illustrating the situation is shown in Fig.~\ref{fig_1}c. Here we explain how to obtain the flux distribution conditional on a spike in one neuron. In the general case of correlated neurons, the action of the full operators $\mathcal{J}_{2D}$ and $\mathcal{D}_{2D}$ is given as
\begin{equation}
	\tilde{P}_{\mathrm{J},kl}=\mathcal{J}_{2D}\mathcal{D}_{2D}P_0=\sum_{k',l'} \sum_{j \: \in \: \mathbb{J}} c(j) (A_j \mathcal{D}_1\otimes B_j \mathcal{D}_2  )_{kl,k'l'} (X^T_{k'} \Omega  Y_{l'}),
\end{equation}			
where $\tilde{P}_{\mathrm{J}}$ is a $(M+m)\times(N+n)$ matrix. The implicitly summed subspace indices, $P_{0,kl}=\sum_{m,n}X^T_{k',m} \Omega_{mn}  Y_{n,l'}$, are not shown and, $A_j$, $B_j$ and $c(j)$ are defined in Eq.~\ref{eq_J_2D}. This equation can be written in a concise form with implicit indices as
\begin{equation*}
	\tilde{P}_{\mathrm{J}}=	\sum_{j \: \in \: \mathbb{J}} c(j)	A_{j}^{T} \mathcal{D}^T_{1}X^T\Omega Y \mathcal{D}_{2} B_{j}=\tilde{X}^{T}\tilde{\Omega} \tilde{Y}.
\end{equation*}
Again, for practical reasons, computations were reduced by projecting $\mathcal{J}_{2D}\mathcal{D}_{2D}$ to extended subspaces $\tilde{X}$ and $\tilde{Y}$ similar to Eq.~\ref{eq_reduced_operators}
\begin{equation}
\tilde{\Omega}  =\sum_{j \: \in \: \mathbb{J}} c(j) \:\Big{[}(\tilde{X}^T A_j X X^T\mathcal{D}_1 X) \otimes (\tilde{Y}^T B_j Y \mathcal{D}_2 Y )\Big{]} \Omega.
\end{equation}
In order to compute the probability of jumps we need to sum the probabilities for a jump over the threshold $V_1>\theta_1$  or $V_2>\theta_2$. The flux distribution is obtained as
\begin{equation}
\label{eq_flux}
P_{\mathrm{flux},k} \propto 	\sum_{l=N}^{N+n}  \tilde{P}_{\mathrm{J},kl}=\sum_{l=N}^{N+n} \tilde{X}^{T}_k\tilde{\Omega}\tilde{Y}_l,
\qquad
P_{\mathrm{flux},l}\propto \sum_{k=M}^{M+m} \tilde{P}_{\mathrm{J},kl}= \sum_{k=M}^{M+m}\tilde{X}^{T}_k\tilde{\Omega}\tilde{Y}_l
\end{equation}
defined for $k<M$ and $l<N$. These expressions are normalized such that $\sum_{k} P_{\mathrm{flux},k} =1$. The amplitude of the delta singularity at the reset potential is obtained as 
\begin{equation}
\label{eq_delta_amp}
r_0=r_{ex,c}\sum_{l=N}^{N+n} \sum_{k=M}^{M+m} \tilde{X}^{T}_k\tilde{\Omega}\tilde{Y}_l ,
\end{equation}
where $r_{ex,c}$ is the rate of excitatory shared spike trains. Once we have found the conditional flux distribution, we can solve the initial value problem defined in Eq.~\ref{eq_IVP} in order to determine the conditional rates $r_{1|2}(t)$ or $r_{2|1}(t)$.

\begin{figure}[h!]
	\begin{center}
		\includegraphics[width=10cm]{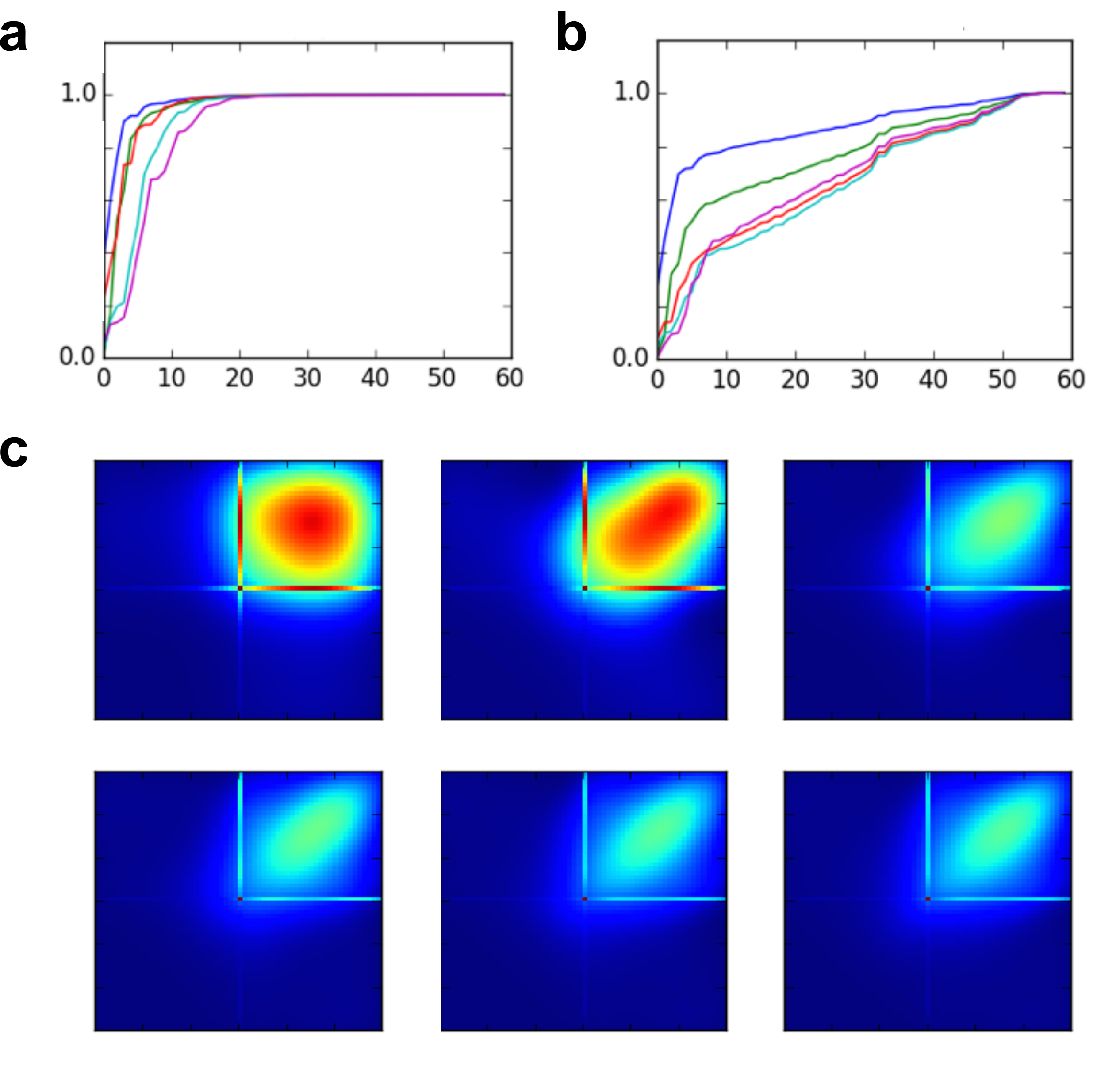}
	\end{center}
	\caption{Reverse engineering of a known stationary distribution. (a)~Projections of a known stationary distribution (obtained by the full jump distribution as in \ref{P_jump_2d}) $P=\mathcal{L}\Sigma \mathcal{R}^T$ on the right singular vectors and (b)~on the left singular vectors of the single-neuron time evolution matrix $\mathcal{M}$. The result is shown here for symmetric neuron parameters $\mathcal{L}=\mathcal{R}=W$. (c)~Reconstruction of a stationary 2D joint membrane potential distribution. Singular vectors sorted by decreasing singular values and added one by one $P_{(K)}=\sum_{k=1}^{K} \sigma_k W_k \otimes W_k$, increasing the number of components from top left to bottom right. The convergence is relatively fast despite the rather high correlation of $c=0.7$. Note that we used here a coarse grid $\Delta V=0.5\,\mathrm{mV}$ as the full solution (vs.~the reduced solution we promote here) of the problem requires $O(N^4)$ operations. }
	\label{fig4}
\end{figure}

\subsection{Diffusion approximation vs.~finite PSPs}

We compare the exit rate of the stochastic system with post-synaptic potentials of finite amplitude with the analytic result obtained for the diffusion approximation \cite{Hakim1999}. For small enough PSPs the difference in rates of the two models is small
\begin{align}
	\label{eq_rates}
	r_{sg}^{-1}&=\tau_{\mathrm{ref}}+\tau_{m}\sqrt{\pi} \int_{\frac{V_{r}-\mu}{\sigma}}^{\frac{V_{th}-\mu}{\sigma}} e^{x^2}[1+\erf(x)] \,dx\\
	r_{\mathrm{Markov}}^{-1}&=\tau_{\mathrm{ref}}+[\frac{1}{h} \sum_{i \geq i_{th}} (\mathcal{J}\mathcal{D}p)_i]^{-1}
\end{align}
We use the absolute difference between the two rates 
\begin{equation}
	r_{\mathrm{error}}(\mu,\sigma)= |r_{sg}(\mu,\sigma)-r_{\mathrm{Markov}}(\mu,\sigma)|
\end{equation}
to account for the accuracy of a specific space-time grid $(\Delta V, \Delta t)$. 

\subsection{Correlation coefficient and comparison to diffusion approximation}

The correlation coefficient in \ref{fig10}a is computed with the formula
\begin{equation}
\label{eq_corr_coeff}
C_{out}(c)=\frac{r_0+\int r_2 (r_{1|2}(\tau)-r_1)\,dt
  +\int r_1 (r_{2|1}(\tau)-r_2)}{CV_1 CV_2\sqrt{r_1r_2}}\,dt. 
\end{equation}
We used \ref{eq_rates} for the stationary rate $r$, $r_0$ is the amplitude of the $\delta$-function  in Eq.~\ref{eq_delta_amp}, and the coefficient of variation,  $CV^2=\frac{\sigma^2_{ISI}}{\mu^2_{ISI}}$, is computed with the equation
\begin{equation}
CV^{2}=2\pi r^2 \int_{\frac{V_{r}-\mu}{\sigma}}^{\frac{V_{th}-\mu}{\sigma}} e^{x^2}\,dx \int^{y}_{-\infty}[1+\erf(x)]^2 \,dy
\end{equation}
as given in \cite{Brunel2000}.
Thereby $\erf(x)$ is the error function \cite{Abramowitz:1974:HMF:1098650}. 
We computed correlation coefficients of the diffusion approximation of the finite PSP system (meaning a 2D Fokker-Planck equation with the same $c$ and $\sigma$s) in \cite{Deniz2016}.

\subsection{Direct numerical simulations and data smoothing}

We used the neural simulation tool NEST \cite{Gewaltig2007} to perform numerical simulations of input and output spike trains in the scenario described above. All analyses were based on discretized voltage data obtained during simulations of $1\,000\,\mathrm{s}$ duration using a time resolution of $\Delta t=10^{-4}\,\mathrm{s}$. 

Empirical voltage distributions were obtained by normalizing histograms appropriately. Further smoothing using a simple moving average was performed before comparing these distributions to the analytically obtained stationary distribution. We also performed the comparison using cumulative distributions, as the implicit integration very efficiently reduces the noise in the data. Two 2D distributions are compared via visual inspection of contour lines. We also directly compare spike train cross-correlation functions to assess efficiency and accuracy of the method. 

\subsection{Numerical evaluation of cross-correlation functions}

We compute cross-correlation functions of spike trains from conditional PSTHs. One can express this as an integral over two variables $\tau=t_1-t_2$ and $s=t_1+t_2$ with bin size $\Delta$
\begin{equation}
C(\tau)=\frac{1}{\Delta}\int_{\tau}^{\tau+\Delta} \frac{d\tau'}{u(\tau')-l(\tau')} 
\int_{l(\tau')}^{u(\tau')} \sum_{i,j} \delta(\tau'-\tau_i)\delta(s'-s_j)\,ds'
\end{equation}
where we set
\begin{equation}
u(\tau)=T/\sqrt{2}+|\tau| ,
\qquad
l(\tau)=T/\sqrt{2}-|\tau|.
\end{equation}
with observation window $T$.
\subsection{Convergence and error bounds}

The direct singular value decomposition of a 2D membrane potential distribution shows that there are only few singular values that deviate significantly from $0$ (Fig.~\ref{fig4}). This behavior does not depend strongly on the chosen discretization, but it does depend on the input correlation coefficient $c$. Although singular vectors are not probability distributions in their own respect, the singular vectors $X_i$ derived from the neuronal dynamics matrix (except the first few vectors) have the property 
\begin{equation}
\sum_k X_{ik} \approx 0
\end{equation}
provided the discretization is fine enough. This behavior is demonstrated in Fig.~\ref{fig6}. The contribution of each sum to the overall normalization
\begin{equation}
\sum_{ij}P_{ij}=\sum_{i,j} \sum_{k=0,l=0}^{K,L} X_{ik} \Omega_{kl} Y_{jl}=\Sigma_1S\Sigma_2
\end{equation}
is progressively small, where $\Sigma_{1k}$ and $\Sigma_{2l}$ are sums of $k$-th and $l$-th singular vectors of the first and the second matrix for $k\geq m$ and $l \geq n$, respectively
\begin{equation}
\label{eq_err_sum}
\Sigma_{err}(m,n)=\sum_{i,j} \sum_{k=m, l=n}^{K,L} X_{ik} \Omega_{kl} Y_{jl}=\sum_{k=m, l=n}^{K,L}  \Sigma_{1k}\Omega_{kl}\Sigma_{2l}.
\end{equation}
This shows that the sum converges rather quickly. This error measure is related to projections of 1D discrete stationary distribution $P_0$ (satisfying $\mathcal{M}_1P_0=P_0$) to SVDs. All other eigenvectors of a Markovian matrix ( $\mathcal{M}_1P_0=\lambda P_0$ for $|\lambda|<1$) satisfy $\sum_k (P_i)_k=0$. We want to avoid underestimating the total probability mass as a result of the truncated sum in Eq.~\ref{eq_p_0}. Hence, above we justify that the remainder of $P_0$  projections after truncation can be omitted up to a certain precision. On the other hand, in order to describe cumulative contribution of singular vectors we look at the $L_1$ distance of the omitted remainder (i.e. $k\geq m$, $l \geq n$)
\begin{equation}
\label{eq_err_sum_2}
E(m,n)=\sum_{i,j}  |\Delta P _{ij}(m,n)|=\sum_{i,j} \Bigl|\sum_{k=m, l=n}^{K,L} X_{ik} S_{kl} Y_{jl} \Bigr|
\end{equation}
which describes how well the method converges self-consistently.  Here we didn't normalize this equation for every term we added. Which means we just rely on fast convergence of $P_0$ projections measured by Eq.~\ref{eq_err_sum}, so first few error terms can be misleading.  


\section{Results}

In order to treat strong correlations we devised a robust numerical method to study the joint statistics of membrane potentials and spike trains of integrate-and-fire model neurons. The case study reported here covers the leaky integrate-and-fire (LIF) model with Poisson input spike trains. However, our method can be easily generalized to non-linear leak functions \cite{GerstnerKistler2002}, conductance based synaptic inputs \cite{Kuhn2004a} and more complex input correlation models \cite{Kuhn2003,Schultze-Kraft2013},. although we have to leave the details of such generalizations open. In this section we will explain how to select a `good basis for expansion', and we will give numerical examples that demonstrate the power of the method. 

\subsection{SVD of joint probability distributions and choice of expansion basis }

We started from a simple observation: The stationary joint membrane potential distribution for two neurons with independent input is given as 
\begin{equation}
P(V_1,V_2)=P_1(V_1)P_2(V_2)
\end{equation}
where $P_1(V_1)$ and $P_2(V_2)$ are the stationary membrane potential distributions of two independent neurons, as described in Eq.~\ref{eq_model}. A similar relation for a discretized voltage grid can be written as
\begin{equation}
P=P_{1} \otimes P_2.
\end{equation}
For the case of shared input this simple relation is not valid any more. On the other hand, we observed that a value of the parameter $c$ close to $0$ will practically recruit only a small number of additional principal components for any given precision, cf.\ 	Fig.~\ref{fig4}. Here we perform a singular value decomposition of the full solution of Eq.~\ref{eq_full_markov} given in terms of the matrix $P_{ij}$
\begin{equation}
P=\mathcal{L}\Sigma \mathcal{R}^{T}
\qquad\text{or}\qquad
P_{ij} = \sum_{k=1}^N \sigma_k L_{ik} R_{kj}
\end{equation}
generalizing the case of independent neurons to also reconstruct the joint membrane potential distribution for neurons with shared input. As demonstrated in Fig.~\ref{fig4}, convergence is rather quick, even for moderate values of $c$.  

Another aim of our study was to gain some understanding about the influence of the space-time grid. We observed that left and right singular vectors are of the form
\begin{equation}
P_{kl}=\sum_{ij} \Omega_{ij} (X^{c}_{ik}+a_i \delta_{k,r})(Y^{c}_{jl}+a_j \delta_{l,r})
\end{equation}
where $X^{c}$ and $Y^{c}$ reflect the quasi-continuous part of basis vectors and $\Omega$ is the coupling matrix as defined in Eq.~\ref{eq_p_0}. Here we need to make sure that the emerging singularity at the reset bin is not causing any numerical problems. One needs to first consider a small time step $\Delta t$ and adapt the stepping in space $\Delta x$ accordingly. A more thorough discussion of a suitable coarse graining strategy, however, is postponed to a later section of this paper. 

\begin{figure}[h!]
	\begin{center}
		\includegraphics[width=12cm]{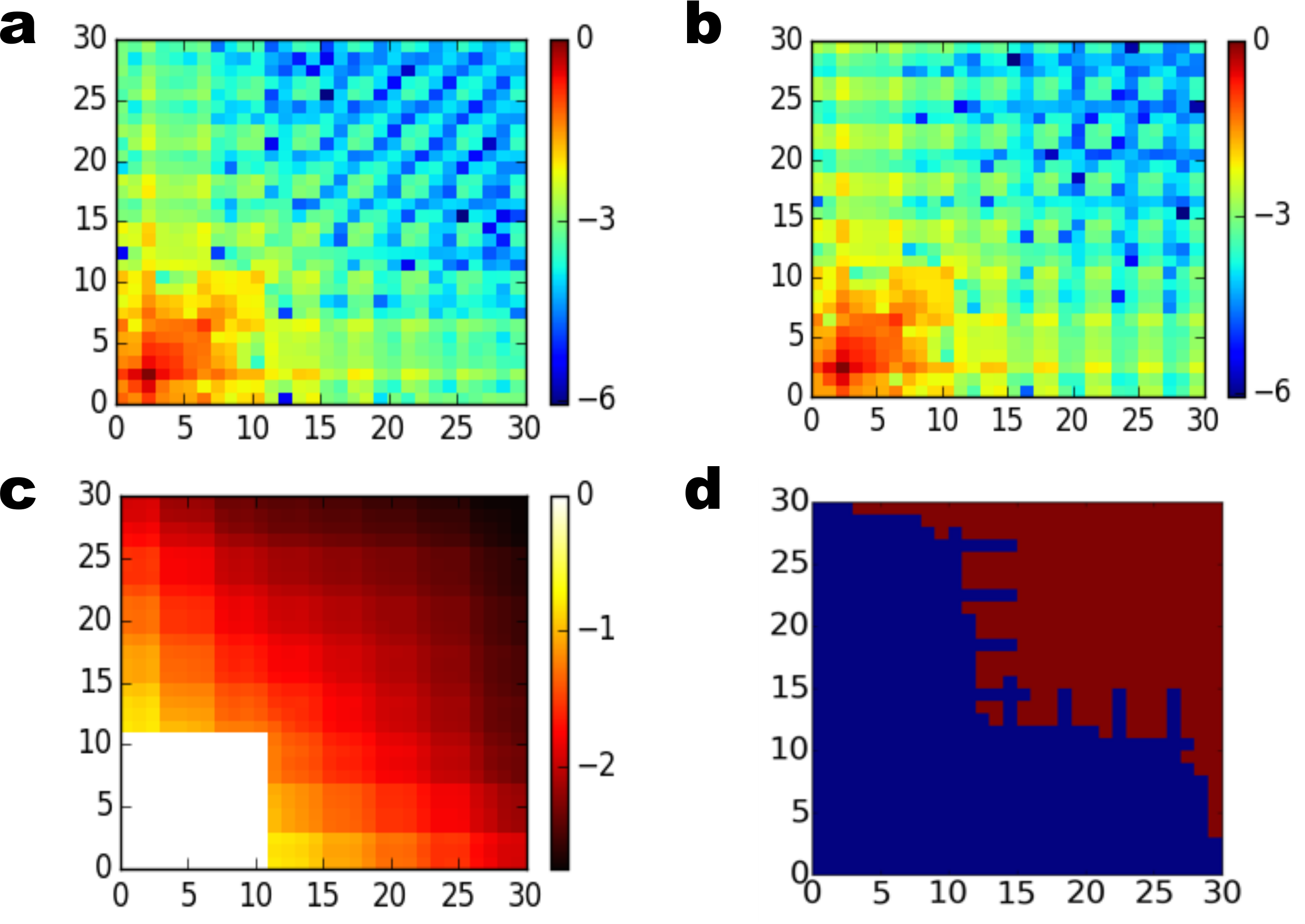}
	\end{center}
	\caption{Comparison of using right or left singular vectors for a reconstruction of the joint membrane potential distribution. We observe that the right singular vectors have better convergence properties. (a)~Mode coupling matrix $\Omega$ for a basis derived from right singular vectors. (c)~Partial sum error (Eq.~\ref{eq_err_sum_2}) for the basis of right singular vectors, corresponding to (a). (b)~Mode coupling matrix $\Omega$ for a basis derived from left singular vectors. (d)~Difference of partial sum errors for left singular vectors corresponding to (b) and right singular vectors corresponding to (a). Red color indicates positive sign, while blue color indicates negative sign of the error. The reconstruction with right singular vectors converges slightly faster. Note that for the error measures considered in (c) and (d) we didn't take into account the bottom left $10\times 10 $ entries of the matrix.}
	\label{fig5}
\end{figure}

\begin{figure}[h!]
	\begin{center}
		\includegraphics[width=12cm]{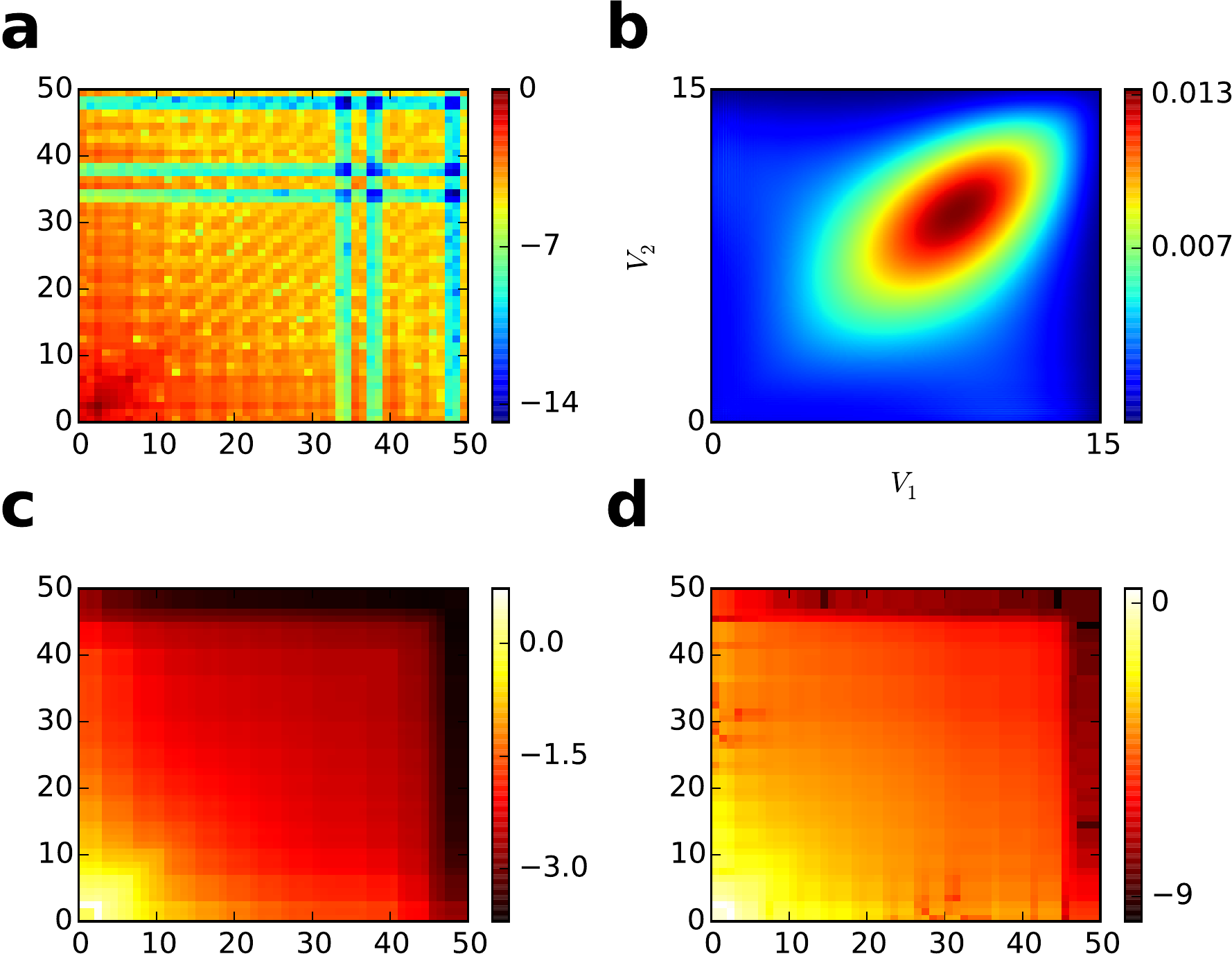}
	\end{center}
	\caption{Convergence of the SVD-based approximation method using up to $50$ singular vectors corresponding to the largest singular values. (a)~Mode coupling matrix $\Omega$, defined by $P_0=X^{T}\Omega Y$ (color represents $\log_{10}(\Omega_{ij}/\max_{ij}(\Omega_{ij}))$). (b)~Reconstructed 2D membrane potential distribution based on a coarse graining with $50 \times 50$ grid points. (c)~$log_{10}$ of relative $L_1$ error. The value given at location $(i,j)$ is the contribution to the reconstruction of $P$ computed via summation of all vectors $n>i$, $m>j$ (Eq.~\ref{eq_err_sum_2}). (d)~Error that arises from $\sum_k X_{ik} \neq 0$ (Eq.~\ref{eq_err_sum}). }
	\label{fig6}
\end{figure}

Here we suggest to use SVD as a method to achieve a dimensional reduction of the full system. As it is a numerical method, its convergence and efficiency needs to be addressed. Generally, there are several different options to select a basis. Specifically, we use the right singular vectors of single-neuron Markov matrices. As demonstrated in Fig.~\ref{fig4}, right singular vectors lead to an expansion that converges faster for coarse grids (e.g.\ $V=0.5\,\mathrm{mV}$). Although for finer grids (e.g.\ $V=0.05\,\mathrm{mV}$) the difference is less prominent (Figs.~\ref{fig5}a and b), right singular vectors still converge slightly faster than left singular vectors (Fig.~\ref{fig5}d). Right singular vectors of the single-neuron time evolution matrix yield an orthogonal coordinate system with very good properties. 

As reported previously \cite{Deniz2016}, we may also use a direct analytical approach using the basis of the single-neuron Fokker-Planck operator, and its adjoint basis
\begin{equation}
P_0(X,Y)=\sum_{ij}\Omega_{ij} X_i \otimes Y_j
\end{equation}
where $X$ and $Y$ are the left eigenvectors of the single neuron operators. However, the issue is that the discrete adjoint basis blows up at the lower boundary. The effect of this on our approximation is demonstrated in Fig.~\ref{fig3}. In general, SVD eliminates a kernel of singular matrices.

In our treatment of the 2D Fokker-Planck equation, which is the infinitesimal limit of the theory considered above, we used the basis and adjoint basis to project linear operators to a subspace. This has certain advantages as it satisfies constraints for marginal distributions and preserves the Markov property to some extent. Positivity of the solution in the subspace is not guaranteed, but time evolution is probability preserving ($\sum_i P_i(t)=1$). 

First, SVD is computationally convenient, because it leads to using a real orthogonal basis which resembles the eigenbasis of $\mathcal{M}$. Second, numerical instabilities due to an ill-conditioned time evolution matrix $\mathcal{M}$ (some eigenvalues $\lambda_i \approx 0$) are cured by SVD. Third, although the basis vectors implied by SVD have the disadvantage of not completely preserving positivity, the deviation remains within tight bounds even for a relatively small number of basis vectors. 


\subsection{Comparison to direct numerical simulations}

\begin{figure}[h!]
	\begin{center}
		\includegraphics[width=12cm]{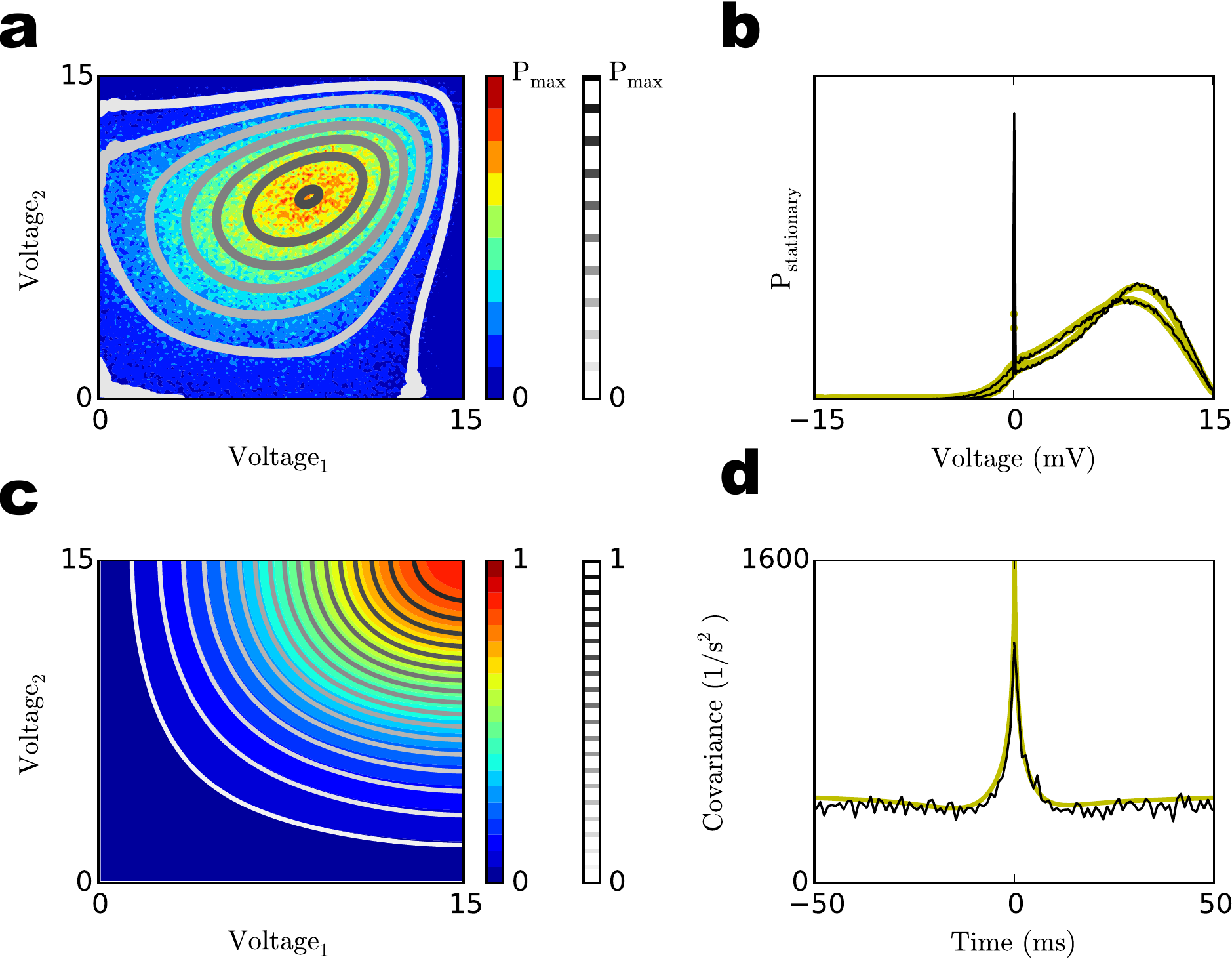}
	\end{center}
	\caption{Effect of different parameters of neurons or input to neurons (here, $\sigma$ asymmetry) on the joint membrane potential distribution and spike cross-covariance function. Reconstruction of the 2D joint membrane potential distribution using SVD ($\Delta V=0.1\,\mathrm{mV}$, $\Delta t=0.1\,\mathrm{ms}$; grey contours) and comparison to direct numerical simulations (color-coded histograms). (a)~Direct comparison of the SVD-based evaluation of the Markov theory and direct simulations. (b)~Comparison of the corresponding 2D cumulative distributions. (c)~Comparison of 1D marginal membrane potential distributions (yellow: Markov theory, black: direct simulation). (d)~Comparison of spike train covariance derived from the Markov theory and direct simulations.  
	}
	\label{fig7}
\end{figure}

We compare our SVD-based Markov theory and direct numerical simulations of spiking neurons both on the level of joint 2D membrane potential distributions and on the level of spike train covariance functions, cf.~Fig.~\ref{fig7}. The empirical distributions derived from 2D histograms are slightly smoothed in order to compare them to the distributions derived from the Markov theory on the level of contour lines. We also considered 2D cumulative distribution functions, where the smoothing step can be omitted. Moreover, we computed output spike train covariance functions as described in methods section and compared them to the covariance functions obtained directly from the simulated spike trains. 

\subsection{Application 1: Non-linear correlation transfer}

Two neurons that are driven by correlated input current will exhibit correlated output spike trains. This transfer of correlation reflects an important property of neuronal dynamics, which is of particular relevance for understanding the contribution of neurons to network dynamics. Recently, we were able to demonstrate, by exact analytical treatment, that the correlation transfer for leaky integrate-and-fire neurons is strongly non-linear \cite{Deniz2016}. Only for weak input correlation it can be described by perturbative methods, and deviations from linear response theory depend on the background firing rate. In the present work we demonstrate the same non-linear correlation transfer, cf.~Fig.~\ref{fig10}. There we also demonstrate how the parameters of the spatial and temporal coarse graining affects the precision of the Markov approximation. Our main conclusion is that dimensional reduction via SVD subspace projections makes it possible to achieve a superior precision with small bin sizes. Fine enough grids, however, could not be dealt with on typical computers without using the reduction suggested here. 

\begin{figure}[h!]
	\begin{center}
		\includegraphics[width=12cm]{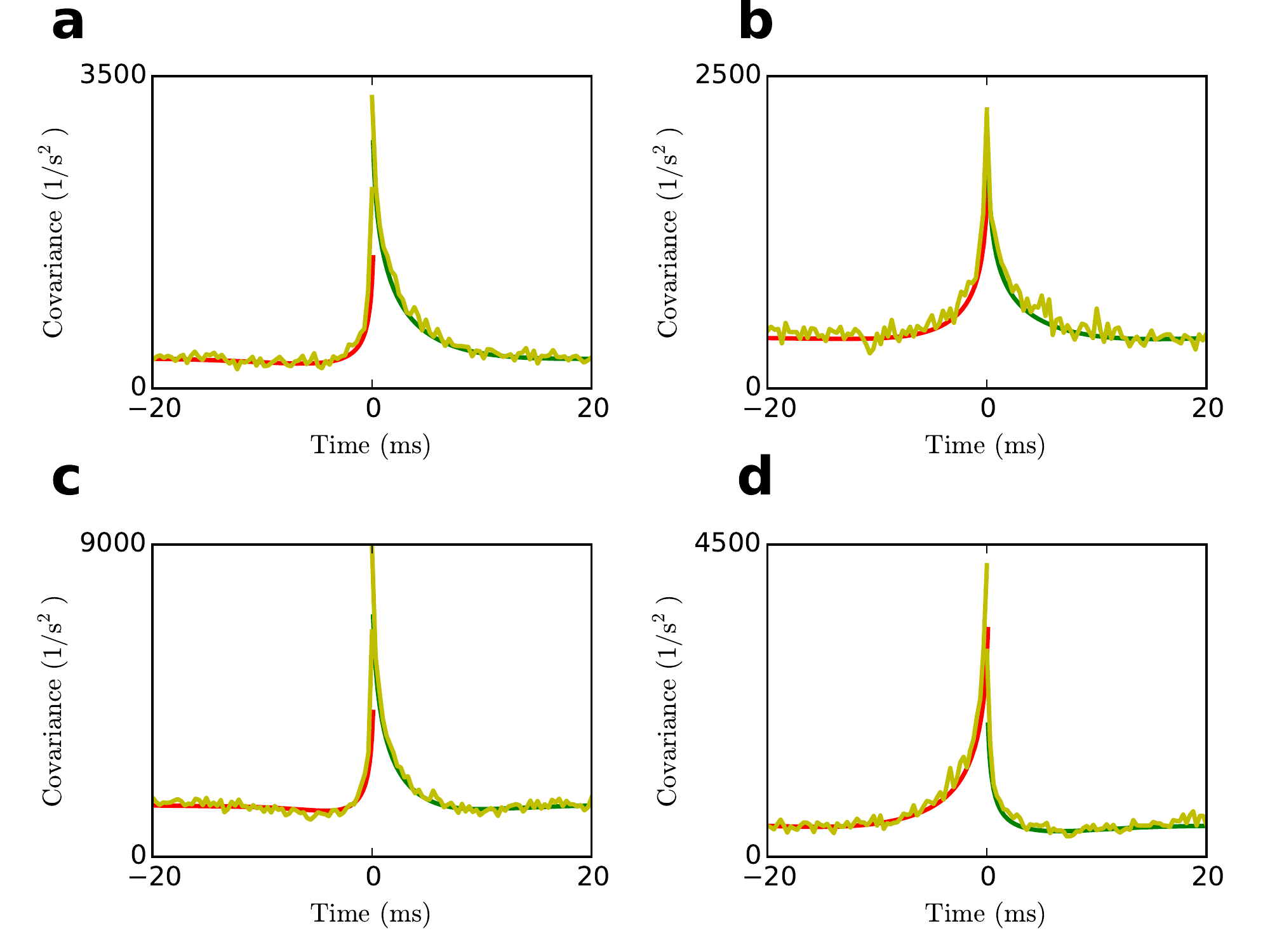}
	\end{center}
	\caption{Asymmetric cross-covariance functions in the strongly correlated regime ($c=0.9$). Covariance functions extracted from simulated spike trains are compared to covariance functions computed with the SVD method suggested in this paper. Results are shown here for different types of asymmetry. (a)~$\mu$ asymmetry, $\mu_1 \neq \mu_2$ while all other paramters are the same, (b)~$\sigma$ asymmetry, $\sigma_1 \neq \sigma_2$ , (c)~$\tau_m$ asymmetry, $\tau_1 \neq \tau_2$, which leads to $\mu_1 \neq \mu_2$ and $\sigma_1 \neq \sigma_2$ as private spike train input rates are the same, (d)~$V_{th}$ asymmetry, $V_{th,1} \neq V_{th,2} $. For specific parameter values, see Table~\ref{Tab:03}.}
	\label{fig8}
\end{figure}

\begin{figure}[h!]
	\begin{center}
		\includegraphics[width=14cm]{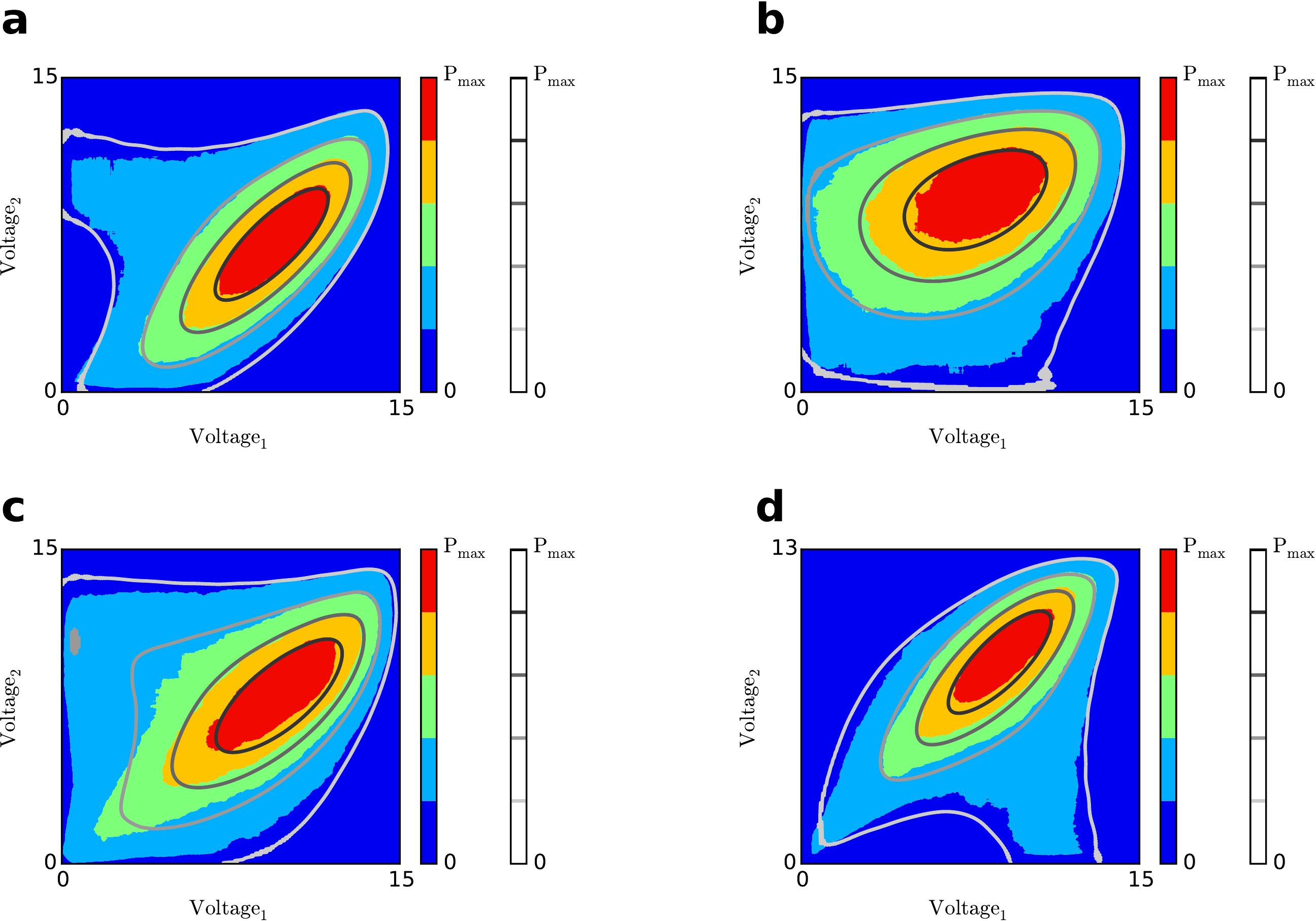}
	\end{center}
	\caption{Joint 2D membrane potential distribution of simulated neuron dynamics for $c=0.9$ (2D histogram smoothed by boxcar kernel of width $w=1\,\mathrm{mV}$) is compared to the joint distribution computed with the SVD method (using a subspace of dimension $50\times 50$). We demonstrate here either heterogeneity in intrinsic parameters, or in input rate $s$. Both types of non-equal neuron parameters can lead to similar distributions. Our method can deal with all such cases accurately. Results are presented here for different asymmetric parameters: (a)~$\mu$ asymmetry, (b)~$\sigma$ asymmetry, (c)~$\tau_m$ asymmetry, which implies an asymmetry in $\mu$ and $\sigma$ as well, as private spike train input rates are the same, (d)~$V_{th}$ asymmetry. For specific parameters, see Table~\ref{Tab:03}.}
	\label{fig9}
\end{figure}


\begin{figure}[h!]
	\begin{center}
		\includegraphics[width=14cm]{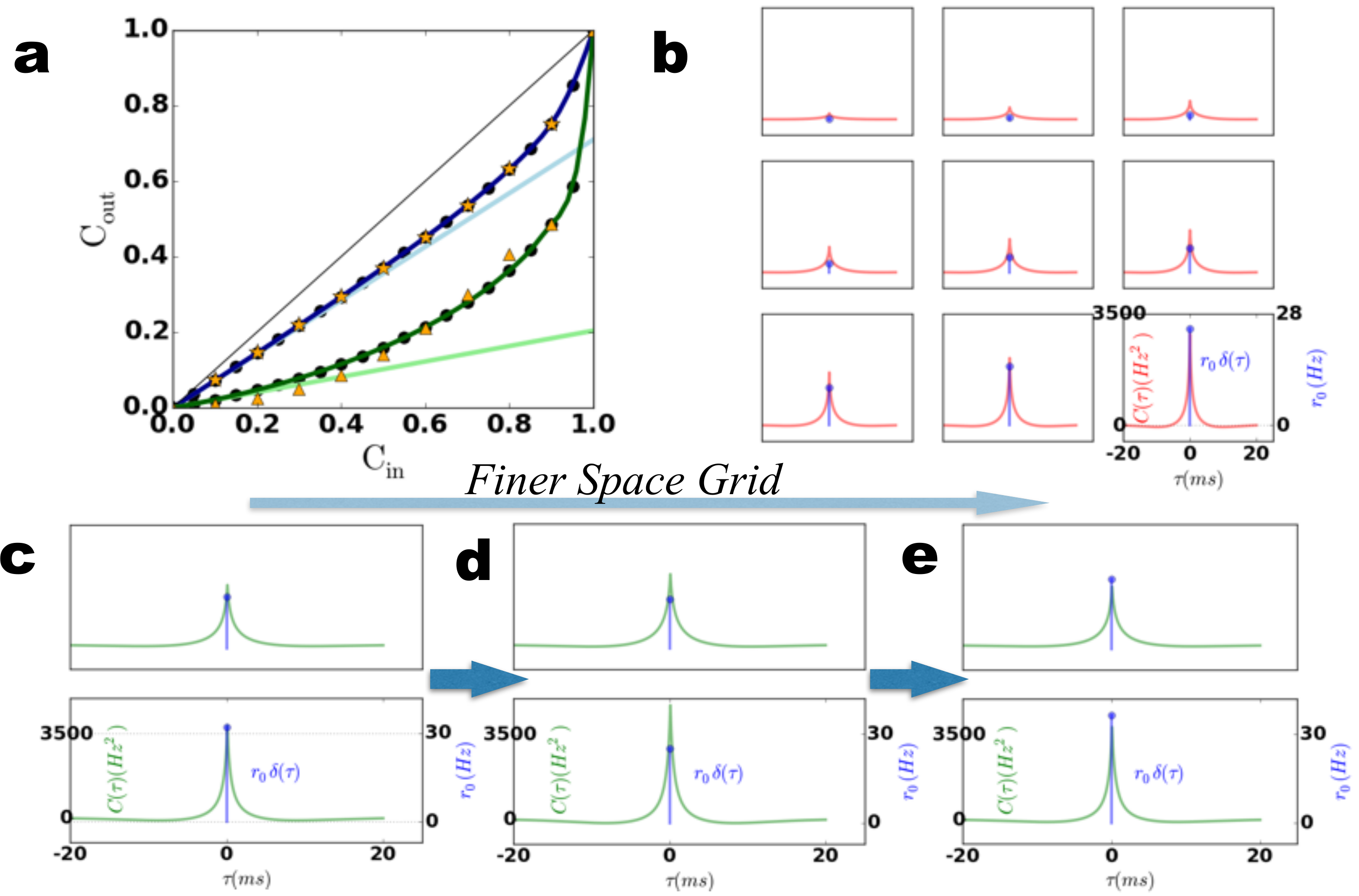} 
	\end{center}
	\caption{Limits to the precision of cross-covariance functions and correlation transfer functions. (a)~Correlation transfer function as a function of input correlation. We compare here analytical results (solid curves) described in a recent paper \cite{Deniz2016} with numerical results (orange symbols) obtained with the methods described in this paper. Non-perturbative correlation transfer functions  $C_{out}(C_{in})$ in \cite{Deniz2016} for symmetric parameters and for high and low firing rates, respectively (blue: $r_b= 15.2\,\mathrm{Hz}$, $\CV^2 =  0.5$; green: $r_g=1.13\,\mathrm{Hz}$, $\CV^2 = 0.98$). Slopes of light blue and light green lines (corresponding to $\frac{dC_{out}}{dC_{in}}$ at $C_{in}=0$) are computed using perturbation theory \cite{DelaRocha2007}. Note that we added the obvious points $C_{out}(0)=0$ and $C_{out}(1)=1$ to the plot by hand. (b)~Cross-covariance functions $C(\tau)$ (solid red curves, with unit ${Hz}^2$) as a function of the lag $\tau$. For non-infinitesimal PSPs there is a delta function at zero lag $\tau=0$ (blue stems, with unit $Hz$), the amplitude of which grows as $c$ increases. Figures from top left to bottom right correspond to different values of $c$. For (a) and (b) we chose $\Delta V=0.05\,\mathrm{mV}$. Panels (c) and (d) are zoomed-in versions of the $c=0.85$ (top) and $c=0.95$ (bottom) covariance functions (solid green curves) to demonstrate the effect of the grid (c)~$\Delta V=0.05\,\mathrm{mV}$ vs.\ (d)~$\Delta V=0.02\,\mathrm{mV}$ vs. (e)~$\Delta V=0.01\,\mathrm{mV}$. 
		Further parameters are given in Table~\ref{Tab:03}}.
	\label{fig10}
\end{figure}

\subsection{Application 2: Asymmetric cross-covariance functions}

Neurons in biological networks have widely distributed parameters, and this heterogeneity may also influence information processing \cite{Padmanabhan2010, Yim2012, Yim2013}. Moreover, robust asymmetries in spike correlations could lead to asymmetric synaptic efficacies, if they are subject to spike timing dependent plasticity \cite{Morrison2008, Babadi2013}. Our approach reveals a generic temporal asymmetry in cross-covariance functions, related to the heterogeneity of intrinsic neuron parameters and input variables. Such temporal asymmetry is more pronounced for larger values of $c$, especially in the non-perturbative
regime that we address in this work.


\begin{figure}[h!]
	\centering
	\includegraphics[width=12cm]{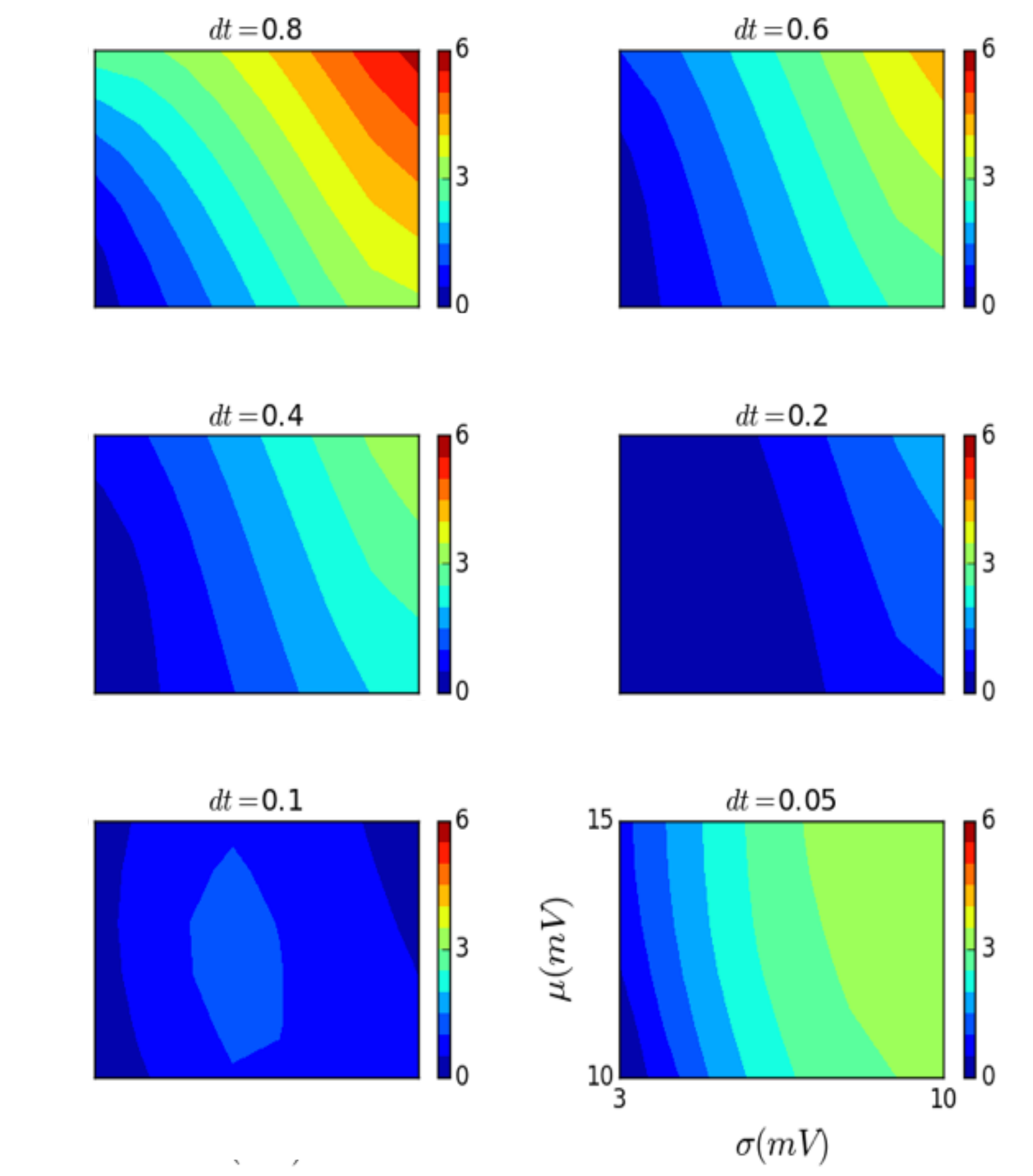}
	\caption{ Here we demonstrate that fixing the space bin (here $\Delta V=0.02\,\mathrm{mV}$), the choice of the time bin affects the firing rate estimation \cite{Helias2010}. 
The deviation of correlation coefficients for small rates in Fig.~\ref{fig10}a (orange triangles vs.\ dark green curve) is a result of a poor estimation of conditional rates.  The variance of the input ($\sigma^2$) is crucial in determining the appropriate temporal bin size, while the mean input ($\mu$) is less effective. With increasing variance one observes an increasing firing rate error. From these plots, we conclude that for a space bin $\Delta V=0.02\,\mathrm{mV}$, a time bin between $\Delta t=0.2\,\mathrm{ms}$ and $\Delta t=0.1\,\mathrm{ms}$ defines a range of good choices. All other parameters are as specified in Table~\ref{Tab:02}. }
		\label{fig11}
	\end{figure}

We document here an application of our method to four types of asymmetries \cite{Yim2013,Deniz2016}: $\mu$ asymmetry, $\sigma$ asymmetry, $\tau_m$ asymmetry and $V_{th}$ asymmetry. We quantified the asymmetry by $a=\chi_1/\chi_2$ (specific values given in Table~\ref{Tab:03}), where $\chi$ is replaced by the respective parameter. Asymmetric correlations have been described previously, and they were by numerical simulations and experimentally studied by \cite{Ostojic2009, Padmanabhan2010, Yim2013, Yim2014}.

***\section{Discussion}

\subsection{Relevance of the new approach presented here}
Models of correlated neuronal activity describe the origin of correlations in spiking model neurons, induced by the structure of the network and/or feedforward input. Such neuron models, however, are notoriously nonlinear. Nevertheless, most treatments rely on linearization and other simplifying assumptions, as nonlinear correlation transfer functions (i.e.\ relation of input and output correlations) are difficult to derive. Previous analytical approaches have employed perturbation theory \cite{Brunel1999, Lindner2001} to study pairwise correlations under the assumption of weak input correlation \cite{DelaRocha2007, Shea-Brown2008}. As a consequence of this technical convenience, we still lack a systematic approach that allows us to deal with a broad range of correlations, and to gain an understanding of their implications for network dynamics. 

\subsection{Extensions of the theory}

All computations described in our paper can be applied to more general integrate-and-fire models with anon-linear leak function  $\Psi(V)$
\begin{equation}
\tau_m\dot{V}=\Psi(V)+J_{ex}\tau_m\sum_k S^{ex}_k(t) +J_{in} \tau_m \sum_k S^{in}_k(t) .
\end{equation}
We only need to rewrite the decay matrix as a discrete approximation of the differential operator $D(x) = \tau_m \frac{d}{dt} x - \Psi(x)$. 

Other scenarios of interest are reflected by an altered amplitude distribution of the inputs. This is a natural consequence if individual synapses have different PSP amplitudes. It could also arise, however, if the population of input neurons has a non-trivial correlation structure. In particular, higher-order correlations have been treated in terms of specific amplitude distributions \cite{Kuhn2003, Staude2010}. The method described in the present paper can be adapted to such scenarios by simply using a modified definition of the jump matrix \cite{Schultze-Kraft2013}.

Higher-order statistics on the output side is also compatible with our method, describing the joint response behavior of three or more neurons that are driven by shared input. Third-order correlations can be computed in practice, because the projections work in the same fashion 
\begin{equation}
P_{lmn}=\sum_{ijk}\Omega_{ijk} X_{il} Y_{jm} Z_{kn}
\end{equation}
now with a 3D jump operator given in the generic form
\begin{equation}
J_{3D}=J_1 \otimes J_2 \otimes J_3.
\end{equation}
This operator is again transformed with a basis derived from a SVD as
\begin{equation}
J_1 \otimes J_2 \otimes J_3 \rightarrow X J_x X^{T} \otimes Y  J_2 Y^{T} \otimes  Z J_3 Z^{T} .
\end{equation}
This procedure is computationally more demanding as we need to consider additional paths, although the scaling is not exponential. 
Under assumption of homogeneous shared input (same jump amplitudes driven by shared input in all directions) leads to an expression similar to Eq.~\ref{eq_J_2D},
\begin{equation}
\label{eq_J_3D}
\mathcal{J}_{3D}=\sum_{j \: \in \: \mathbb{J}} \psi(j)[e^{J_1}O^{j}\otimes e^{J_2}O^{j} \otimes e^{J_2}O^{j} ].
\end{equation} 
Assuming joint stationarity of all three spike trains $S_1(t)$, $S_2(t)$ and $S_3(t)$, we need to find the joint third moment of the spike train statistics
\begin{equation}
\label{eq_third_moment}
\mu_{123} (\tau_1,\tau_2)= \bigl\langle S_1(\tau_1) S_2(\tau_2) S_3(0) \bigr\rangle.
\end{equation}
As shown above, second moments can be computed with our method (Fig.~\ref{fig9}). In order to obtain the covariance function from the stationary 3D flux, the time evolution of the 2D conditional flux at times $(\tau_1, \tau_2)$ is needed. This is given as
\begin{equation}
\Delta_t P_{12|3}(t)=\mathcal{M}_{12} P_{12|3}(t)
\end{equation} 
which is computationally demanding as the numerical effort scales as $O(N^6)$. However, this can be projected to the subspace with time dependent coupling matrix $\Omega$ as
\begin{equation}
\label{eq_2D_reduced}
\Delta_t \Omega(t)=\mathcal{Q}_{12} \Omega(t).
\end{equation} 
This form has advantages over finite difference methods as e.g.\ suggested in \cite{Rosenbaum2012}. 
The computation of the third order moment defined in Eq.~\ref{eq_third_moment} requires a solution of Eq.~\ref{eq_2D_reduced} at $\tau_2$ to find the second conditional distribution $P_{1|2|3}(\tau_2)$. Then we need to find the 1D conditional distribution (e.g.\ for neuron 1) at $t=\tau_1$ by solving 
\begin{equation}
\Delta_t P_{1|2|3}(t)=\mathcal{M}_{1} P_{1|2|3}(t)
\end{equation} 
similar to Eq.~\ref{eq_IVP}. This provides us with conditional rate $r_{1|2|3}(\tau_1,\tau_2)$ and then the third moment is given as
\begin{equation}
\mu_{123} (\tau_1,\tau_2)=r_{1|2|3}(\tau_1,\tau_2)r_3.
\end{equation}
Details of this computation have to be deferred to future work, though. 

\subsection{Boundary conditions and singularity}

The joint membrane potential distribution has a singularity at the origin $(V_{1}, V_{2})=(0,0)$ due to a coordinated reset caused by some shared input spikes. There is also a line discontinuity at $V_1=0$ and $V_2=0$, again due to the reset boundary condition. These singularities are reflected in the right singular vectors $X$ and $Y$. This is the exact reason why we selected them as a basis to expand operators and joint distributions.

We observed that a singularity (a $\delta$-function) emerges when $\Delta V$ is small and $\Delta t$ is large, in relative terms. This is an issue even for the 1D discrete problem, and it is even more severe for 2D problems as the amplitude of the singularity scales quadratically with $\Delta V$. This phenomenon occurs only if PSPs have a finite amplitude. As the PSP gets larger relative to $\Delta V$, reset currents remain finite even in continuous time \cite{Helias2010}. As a consequence, the limit to continuous variables must be taken with care, in particular for $c>0$. 

The $\delta$-singularity does not exist for the diffusion approximation \cite{Deniz2016}. However, the definition of the current at the origin again fails as the derivative is discontinuous in both $V_1$ and $V_2$ directions. The infinitesimal limit of the jump equation must be taken with care. There is no doubt that the jump equation is well-defined as the flux at the boundary is not local. However, the infinitesimal limit is problematic for correlated neurons ($c>0$) as the flux is not defined at the boundary of the 2D domain.

\subsection{Precision, computational efficiency and grid selection}

The selection of an appropriate grid in space and time is crucial for correlation computations. The small residual offset between direct simulation and our new semi-analytical computation (cf.\ Fig.~\ref{fig8} and \ref{fig9}), for example, can be considered as a discretization artifact. Although this issue would deserve a more systematic treatment, we report here some observations that can guide grid selection: 
\begin{enumerate}
\item[(i)] For discrete solutions of the heat equation based on central difference scheme, convergence of 1D time evolution requires $\frac{\Delta t}{(\Delta V)^2} < \frac{\tau}{\sigma^2}$ \cite{Ames2014}. A similar rule also applies in the 2D case considered here. In general, explicit discretization schemes of second order differential operators arising in the study of diffusion, require positivity and stability conditions in the order of $\Delta t = O(\frac{\tau (\Delta V_1+\Delta V_2)^2}{\sigma^2})$ \cite{Rosenbaum2012}.
In this work, we followed a discretization scheme that approximately conserves probability, a Markovian approximation \cite{Helias2010}. However, we note once more that some grids may lead to violation of the Markov property for too large $\Delta t$, as a result of boundary effects. This may create issues when the largest eigenvalue exceeds $1$.
 
\item[(ii)] To reflect small expected bin counts (especially for $c \approx 1$) adequately, one needs a larger $\Delta t$ and a smaller $\Delta V$. This is in conflict with rule (i). 
Besides, we observe that smaller $\Delta V$ for a fixed $\Delta t$ actually leads to better firing rate approximation up to some point (Fig.~\ref{fig11}).

\item[(iii)] A finer grid requires more computational efforts to achieve a smooth correlation function. The SVD reduction does not alter this behavior. Other dependencies and limiting factors are indicated in Fig.~\ref{fig10}.  There are two constraining factors which are determined by the selected precision of the approximation. One is the extent of the jump distribution, which affects the number of terms to be accumulated (size of the set  $\mathbb{J}$ in Eq.~\ref{eq_J_2D}). For a fixed grid and a selected precision, this number increases with $\sigma_c$. The second constraint is the size of the SVD subspace. We know that as $c$ gets closer to $1$ and $C(\tau)$ gets steeper we need to include more singular components.
\end{enumerate}

In Fig.~\ref{fig10} we illustrate how coarse graining affects the shape of the cross-covariance function $C(\tau)$. Although the precision of the approximation is limited by the subspace projections implied by SVD, the grid parameters $\Delta t$ and $\Delta V$ are the most important factors to get the shape of the function right. However, for a fixed dimension of the SVD subspace even the finest grid would not be able to capture the singularity at zero time lag ($\tau=0$). The grid effectively limits the precision of the approximation due to the reduced number of degrees of freedom.


\section{Conclusion}

We developed a novel numerical method to compute the joint statistics and correlation functions for two LIF model neurons driven by shared input. Our approach can deal with the full range of input correlations $c$, and the expansion converges fast. Also, our method is widely generalizable and can deal with other scenarios that are biologically relevant. We observed in previous work \cite{Deniz2016} that low output firing rates generally require a non-perturbative treatment. If output rates are high, in contrast, and for high input firing rates with small PSPs (diffusion regime), the approximation derived from linear response theory \cite{DelaRocha2007} is reasonably precise.

We conclude that it is possible to compute correlation functions (in contrast to deriving them from simulations) for a wide range of models with finite PSP amplitudes, and also for a wide range of parallel spike train input models. Although there is currently no conclusive theory for the selection of an appropriate spatio-temporal grid, we were able to come up with some heuristics. The precision of even the first moment (firing rate) depends on the grid. Specifically if $c$ is close to $1$, the temporal component of the correlation function resembles a delta function. In order to capture this phenomenon the grid must be fine enough.

The innovation in our work is not only the formulation of correlation functions based on a Markov chain approximation, but also a dimensional reduction. This helps us compute joint membrane potential distributions. We showed that the number of components obtained by SVD needed to represent single neuron dynamical evolution matrices is small. This also means that computations can presumably be generalized to higher-order correlations with only moderately increased computational effort. 

Systematic benchmarking of our method has not yet been performed. However, we believe our method constitutes the only reasonable numerical approximation to the joint statistics of strongly correlated neuronal dynamics with finite PSPs, apart from direct stochastic simulations \cite{Richardson2010}. This approximation for reasonable grids in space and time was only viable with SVD subspace projections.

\newpage 
\section*{Appendix: Parameters}

%
%
%
%
%
%

\begin{center}
		\begin{table}[h!]
			\textbf{\refstepcounter{table}\label{Tab:02} Table \arabic{table}.}{ Parameters for NEST simulations and semi-analytical computations }
			
			{\begin{tabular}{llr}
					
					\hline
					
					\multicolumn{2}{c}{\textbf{ Neuron parameters: ( Fig.~\ref{fig3}, Fig.~\ref{fig5}, Fig.~\ref{fig6}, Fig.~\ref{fig7})}} \\
					\hline
					Symbol    & Description & Value  \\
					\hline
					$V_{th}$       & voltage threshold     & 15 mV      \\
					$V_{r}$      & voltage reset    & 0 mV     \\
					$\tau_m$ & membrane time constant      & 15 ms       \\
					$\tau_{ref} $ & refractory period   & 1 ms   \\
					$h$ &  PSP      & 0.01 mV -0.1 mV      \\
					$\Delta t$ & time resolution & 0.1 ms \\
					\hline
					
					\multicolumn{2}{c}{\textbf{ Input parameters}} \\
					\hline
					
					$\mu$       & mean input   & 10-15 mV      \\
					$\sigma_1$ ,  $\sigma_2$     & STD private input    & 2-10 mV     \\
					$\sigma_c$       &  STD shared input    & 2-10 mV      \\
					$c $      & input noise correlations    & 0-1    \\
					$a $       &  asymmetry factor  & $>0$\\
					
				\end{tabular}}{}
			\end{table}
			
		\end{center}
		

		\begin{center}
			
			\begin{table}[h!]
				\textbf{\refstepcounter{table}\label{Tab:03} Table \arabic{table}.}{ Numerical results vs.\  NEST simulations : (Fig.~\ref{fig8}, Fig.~\ref{fig9}, Fig.~\ref{fig10})\\ }
				{\begin{tabular}{llr}
						
						\hline			
						\multicolumn{2}{c}{\textbf{ Correlation asymmetry parameters }} \\
						\hline
						Symbol    & Description & Value  \\
						\hline
						$V_{th}$       & voltage threshold     & 15 mV      \\
						$V_{r}$      & voltage reset    & 0 mV     \\
						$\tau_m$ & membrane time constant      & 15 ms       \\
						$\tau_{ref} $ & refractory period   & 1 ms   \\
						$h$ &  PSP      & 0.1 mV      \\
						$\mu$       & mean input   & 12. mV      \\
						$\sigma_0$       &  STD total input    & 5. mV for  (b) and (d) , 6. mV for  (a) and (c)   \\
						$c $      & input noise correlations    & 0.9    \\
						\hline
						$a_{\sigma} $       &  asymmetry factor         &  $1/\sqrt{2}$  \\
						$a_{\mu} $       &  asymmetry factor        & 10/13  \\
						$a_{\tau} $       &  asymmetry factor        & 10/15  \\
						$a_{V_{th}} $       &  asymmetry factor        & 13/15  \\
						\hline
						&*** asymmetry factors:  $a=\chi_1/\chi_2$.
					\end{tabular}}{}
				\end{table}
			\end{center}
			
\clearpage

\begin{center}
\begin{table}[h!]
\textbf{\refstepcounter{table}\label{Tab:04} Table \arabic{table}}{ \textbf{Neuron model parameters}  \\ }
{\begin{tabular}{llr}
	
	\hline			
	Symbol    & Description & Unit \\
	\hline
	$S ^{ex}$ ,$S ^{in}$      & spike trains    &  1/ms    \\
	$V$       & membrane potential    &  mV      \\
	$V_{th}$       & voltage threshold     &  mV      \\
	$V_{r}$      & voltage reset    &  mV     \\
	$t$				& time 					& ms \\
	$\tau_{m,1}$ ,$\tau_{m,2}$ & membrane time constant      & ms       \\
	$\tau_{ref} $ & refractory period   &  ms   \\
	$h_{ex}=h$, $h_{in}=gh$ &  PSP      & mV      \\
	$\mu$       & mean input   &  mV      \\
	$\sigma_1$,$\sigma_2$       &  STD private input    &  mV      \\
	$\sigma_c$       &  STD shared input    &  mV      \\
	$\sigma_0$       &  STD total input    &  mV      \\
	$c $      & input noise correlations    & 0-1   \\
	
	$r_e$, $r_i$ & ex \& inh input rates          & Hz \\
	$r_1$, $r_2$ & output rates of 2 neruons          & Hz \\
		\hline
		\end{tabular}}{}
	\end{table}
\end{center}

\begin{center}
	\begin{table}[h!]
		\textbf{\refstepcounter{table}\label{Tab:05} Table \arabic{table}}{ \textbf{Correlations and related notation}  \\ }
		{\begin{tabular}{llr}
				
				\hline			
				Symbol    & Description & Unit \\
				\hline		
		$C (\tau)$     & covariance function   &  $\text{Hz}^2$   \\
		$r_{1|2}(\tau)$       & conditional rate   &  $\text{Hz}$      \\
		$P(V)$       & probability distribution of V    &  1/mV      \\
		$P_{1|2}(V_1)$       & conditional probability distribution of $V_1$    &  $1/\text{mV}$      \\
		$P(V_1,V_2)$       & joint probability distribution of $(V_1,V_2)$    &  $1/\text{mV} ^2$     \\
		$\Delta_t $				& discrete time evolution operator				& 1 \\
		$\mathcal{M}_1 $				& discrete time evolution matrix & 1\\
		\hline\\
	\end{tabular}}{}
\end{table}
\end{center}
	
		
\begin{center}
	\begin{table}[h!]
		\textbf{\refstepcounter{table}\label{Tab:06} Table \arabic{table}}{ \textbf{Probability distributions and Markov approximation}  \\ }
		{\begin{tabular}{llr}
				
				\hline			
				Symbol    & Description & Unit \\
				\hline		
			$P(q,r)$       & probability for $q$ excitatory, $r$ inhibitory spike     &  $1$     \\
			$P(\gamma)$ & probability for jumps of length $\gamma$     & 1 \\
			$P(V_1,V_2|V_{1,0},V_{2,0})$       & probability forjumps from $(V_{1,0},V_{2,0})$ to  $(V_1,V_2)$    &  $1$     \\
			$P(V_1,V_2)$       & probability for jumps from $(0,0)$ to  $(V_1,V_2)$    &  $1$     \\
			$\mathcal{T}$     & threshold matrix : \small{$N\times (N+n)$}&  1   \\
			$\mathcal{J}$     & jump matrix  : \small{$(N+n)\times N$ }&  1  \\
			$\mathcal{D}$     & decay matrix: \small{$N\times N$ }& 1 \\
			$\mathcal{M}$     & time evolution matrix :\small{$N\times N$ }& 1 \\
			$\mathcal{T}_{2D}$     & threshold tensor : \small{$N\times M\times (N+n)\times (M+m) $}&  1   \\
			$\mathcal{J}_{2D}$     & jump tensor  : \small{$(N+n)\times (M+m) \times N\times M$ }&  1  \\
			$\mathcal{D}_{2D}$     & decay tensor: \small{$N\times M\times N\times M$} & 1 \\
			$\mathcal{M}_{2D}$     & Time evolution tensor 2D :\small{$N\times M\times N\times M$} & 1 \\
			\hline
			$U$,$D$				& up or down operators in discrete space  & 1 \\	
			$J^p_{1}$, $J^p_{2}$ & private $V_1$ or $V_2$ jump generators  & 1 \\	
			$J^c_{1}$, $J^c_{2}$ & shared $V_1$ or $V_2$ jump generators& 1 \\	
			$c_{mn}$    & coefficient of $m$ excitatory and $n$ inhibitory jumps& 1 \\
			\hline
			$P_0$	    & stationary probability density in discrete space    &  $(\text{mV})^{-2}$  \\ 
			$P_{J,k}$       & probability for jumps $k$-th component of $\mathcal{J}P_0$   &  $1/\text{mV}$    \\
			$P_{flux,k}$       & probability for jumps $k$-th component conditional exit flux   &  $1/\text{mV}$    \\
			$\Delta t$    & time bin & ms \\
			$\Delta V$    & voltage  bin & mV \\
			$a_1$ , $b_1$     & average count of private input $1$ in $\Delta t$   & 1 \\
			$a_2$ , $b_2$     & average count of private input $2$ in $\Delta t$   & 1 \\
			$a_c$ , $b_c$     & average count of shared input in $\Delta t$   & 1 \\				
				\hline
		\end{tabular}}{}
	\end{table}
\end{center}
					
\begin{center}
	\begin{table}[h!]
		\textbf{\refstepcounter{table}\label{Tab:07} Table \arabic{table}}{ \textbf{SVD reduction}  \\ }
		{\begin{tabular}{llr}
				
				\hline			
				Symbol    & Description & Unit \\
				\hline		
		$\mathcal{L}$, $\mathcal{R}$      & SVD left and right basis  &  1   \\
				$\Sigma$      & singular value matrix &  1   \\
		$X$, $Y$      & SVD subspaces    &  1   \\
		$\tilde{X}$, $\tilde{Y}$      & extended subspace    &  1   \\
		$\mathcal{Q}$ & reduced operators defined on a selected subspace    &  1   \\
		$M$ or $N$    & size of full grid i.e.\ matrix & 1\\
		$m$ or $n$    & maximum number of jumps over the threshold & 1\\
		$K$ or $L$    & size of SVD subspace & 1\\
				$M+m$ or $N+n$    & size of full jump subspace & 1\\
				$K+m$ or $L+n$    & size of reduced jump subspace & 1\\
		\hline
	\end{tabular}}{}
\end{table}
\end{center}
				
\begin{center}
	\begin{table}[h!]
		\textbf{\refstepcounter{table}\label{Tab:08} Table \arabic{table}}{ \textbf{Asymmetry parameters}  \\ }
		{\begin{tabular}{llr}				
				\hline			
				Symbol    & Description & Unit \\
				\hline
				$a_{\sigma} $       &  asymmetry factor  $\sigma_1/\sigma_2$       &  1 \\
				$a_{\mu} $       &  asymmetry factor    $\mu_1/\mu_2$     & 1\\
				$a_{\tau} $       &  asymmetry factor    $\tau_1/\tau_2$     & 1 \\
				$a_{V_{th}} $       &  asymmetry factor    $V_{th,1}/V_{th,2}$     & 1 \\
				\hline
			\end{tabular}}{}
		\end{table}
	\end{center}

\newpage
\bibliographystyle{apsrev4-1}
\bibliography{BIBLIO} 			

\begin{thebibliography}{47}%
\makeatletter
\providecommand \@ifxundefined [1]{%
 \@ifx{#1\undefined}
}%
\providecommand \@ifnum [1]{%
 \ifnum #1\expandafter \@firstoftwo
 \else \expandafter \@secondoftwo
 \fi
}%
\providecommand \@ifx [1]{%
 \ifx #1\expandafter \@firstoftwo
 \else \expandafter \@secondoftwo
 \fi
}%
\providecommand \natexlab [1]{#1}%
\providecommand \enquote  [1]{``#1''}%
\providecommand \bibnamefont  [1]{#1}%
\providecommand \bibfnamefont [1]{#1}%
\providecommand \citenamefont [1]{#1}%
\providecommand \href@noop [0]{\@secondoftwo}%
\providecommand \href [0]{\begingroup \@sanitize@url \@href}%
\providecommand \@href[1]{\@@startlink{#1}\@@href}%
\providecommand \@@href[1]{\endgroup#1\@@endlink}%
\providecommand \@sanitize@url [0]{\catcode `\\12\catcode `\$12\catcode
  `\&12\catcode `\#12\catcode `\^12\catcode `\_12\catcode `\%12\relax}%
\providecommand \@@startlink[1]{}%
\providecommand \@@endlink[0]{}%
\providecommand \url  [0]{\begingroup\@sanitize@url \@url }%
\providecommand \@url [1]{\endgroup\@href {#1}{\urlprefix }}%
\providecommand \urlprefix  [0]{URL }%
\providecommand \Eprint [0]{\href }%
\providecommand \doibase [0]{http://dx.doi.org/}%
\providecommand \selectlanguage [0]{\@gobble}%
\providecommand \bibinfo  [0]{\@secondoftwo}%
\providecommand \bibfield  [0]{\@secondoftwo}%
\providecommand \translation [1]{[#1]}%
\providecommand \BibitemOpen [0]{}%
\providecommand \bibitemStop [0]{}%
\providecommand \bibitemNoStop [0]{.\EOS\space}%
\providecommand \EOS [0]{\spacefactor3000\relax}%
\providecommand \BibitemShut  [1]{\csname bibitem#1\endcsname}%
\let\auto@bib@innerbib\@empty
\bibitem [{\citenamefont {Rosenbaum}\ \emph {et~al.}(2014)\citenamefont
  {Rosenbaum}, \citenamefont {Tchumatchenko},\ and\ \citenamefont
  {Moreno-Bote}}]{Rosenbaum2014}%
  \BibitemOpen
  \bibfield  {author} {\bibinfo {author} {\bibfnamefont {R.}~\bibnamefont
  {Rosenbaum}}, \bibinfo {author} {\bibfnamefont {T.}~\bibnamefont
  {Tchumatchenko}}, \ and\ \bibinfo {author} {\bibfnamefont {R.~A.}\
  \bibnamefont {Moreno-Bote}},\ }\href {\doibase 10.3389/fncom.2014.00102}
  {\emph {\bibinfo {title} {Frontiers in Computational Neuroscience}}},\
  Vol.~\bibinfo {volume} {8}\ (\bibinfo {year} {2014})\BibitemShut {NoStop}%
\bibitem [{\citenamefont {Lampl}\ \emph {et~al.}(1999)\citenamefont {Lampl},
  \citenamefont {Reichova},\ and\ \citenamefont {Ferster}}]{Lampl1999}%
  \BibitemOpen
  \bibfield  {author} {\bibinfo {author} {\bibfnamefont {I.}~\bibnamefont
  {Lampl}}, \bibinfo {author} {\bibfnamefont {I.}~\bibnamefont {Reichova}}, \
  and\ \bibinfo {author} {\bibfnamefont {D.}~\bibnamefont {Ferster}},\
  }\href@noop {} {\bibfield  {journal} {\bibinfo  {journal} {Neuron}\ }\textbf
  {\bibinfo {volume} {22}},\ \bibinfo {pages} {361} (\bibinfo {year}
  {1999})}\BibitemShut {NoStop}%
\bibitem [{\citenamefont {Okun}\ and\ \citenamefont {Lampl}(2008)}]{Okun2008}%
  \BibitemOpen
  \bibfield  {author} {\bibinfo {author} {\bibfnamefont {M.}~\bibnamefont
  {Okun}}\ and\ \bibinfo {author} {\bibfnamefont {I.}~\bibnamefont {Lampl}},\
  }\href {\doibase 10.1038/nn.2105} {\bibfield  {journal} {\bibinfo  {journal}
  {Nature Neuroscience}\ }\textbf {\bibinfo {volume} {11}},\ \bibinfo {pages}
  {535} (\bibinfo {year} {2008})}\BibitemShut {NoStop}%
\bibitem [{\citenamefont {Poulet}\ and\ \citenamefont
  {Petersen}(2008)}]{Poulet2008}%
  \BibitemOpen
  \bibfield  {author} {\bibinfo {author} {\bibfnamefont {J.~F.}\ \bibnamefont
  {Poulet}}\ and\ \bibinfo {author} {\bibfnamefont {C.~C.}\ \bibnamefont
  {Petersen}},\ }\href@noop {} {\bibfield  {journal} {\bibinfo  {journal}
  {Nature}\ }\textbf {\bibinfo {volume} {454}},\ \bibinfo {pages} {881}
  (\bibinfo {year} {2008})}\BibitemShut {NoStop}%
\bibitem [{\citenamefont {Arieli}\ \emph {et~al.}(1996)\citenamefont {Arieli},
  \citenamefont {Sterkin}, \citenamefont {Grinvald},\ and\ \citenamefont
  {Aertsen}}]{Arieli1996}%
  \BibitemOpen
  \bibfield  {author} {\bibinfo {author} {\bibfnamefont {A.}~\bibnamefont
  {Arieli}}, \bibinfo {author} {\bibfnamefont {A.}~\bibnamefont {Sterkin}},
  \bibinfo {author} {\bibfnamefont {A.}~\bibnamefont {Grinvald}}, \ and\
  \bibinfo {author} {\bibfnamefont {A.}~\bibnamefont {Aertsen}},\ }\href
  {\doibase 10.1126/science.273.5283.1868} {\bibfield  {journal} {\bibinfo
  {journal} {Science (New York, N.Y.)}\ }\textbf {\bibinfo {volume} {273}},\
  \bibinfo {pages} {1868} (\bibinfo {year} {1996})},\ \Eprint
  {http://arxiv.org/abs/arXiv:1011.1669v3} {arXiv:arXiv:1011.1669v3}
  \BibitemShut {NoStop}%
\bibitem [{\citenamefont {Staude}\ \emph {et~al.}(2008)\citenamefont {Staude},
  \citenamefont {Rotter},\ and\ \citenamefont {Gr{\"u}n}}]{Staude2008}%
  \BibitemOpen
  \bibfield  {author} {\bibinfo {author} {\bibfnamefont {B.}~\bibnamefont
  {Staude}}, \bibinfo {author} {\bibfnamefont {S.}~\bibnamefont {Rotter}}, \
  and\ \bibinfo {author} {\bibfnamefont {S.}~\bibnamefont {Gr{\"u}n}},\
  }\href@noop {} {\bibfield  {journal} {\bibinfo  {journal} {Neural
  Computation}\ }\textbf {\bibinfo {volume} {20}},\ \bibinfo {pages} {1973}
  (\bibinfo {year} {2008})}\BibitemShut {NoStop}%
\bibitem [{\citenamefont {Kumar}\ \emph {et~al.}(2010)\citenamefont {Kumar},
  \citenamefont {Rotter},\ and\ \citenamefont {Aertsen}}]{Kumar2010}%
  \BibitemOpen
  \bibfield  {author} {\bibinfo {author} {\bibfnamefont {A.}~\bibnamefont
  {Kumar}}, \bibinfo {author} {\bibfnamefont {S.}~\bibnamefont {Rotter}}, \
  and\ \bibinfo {author} {\bibfnamefont {A.}~\bibnamefont {Aertsen}},\ }\href
  {http://dx.doi.org/10.1038/nrn2886} {\bibfield  {journal} {\bibinfo
  {journal} {Nat Rev Neurosci}\ }\textbf {\bibinfo {volume} {11}},\ \bibinfo
  {pages} {615} (\bibinfo {year} {2010})}\BibitemShut {NoStop}%
\bibitem [{\citenamefont {Doiron}\ \emph {et~al.}(2016)\citenamefont {Doiron},
  \citenamefont {Litwin-Kumar}, \citenamefont {Rosenbaum}, \citenamefont
  {Ocker},\ and\ \citenamefont {Josi{\'{c}}}}]{Doiron2016a}%
  \BibitemOpen
  \bibfield  {author} {\bibinfo {author} {\bibfnamefont {B.}~\bibnamefont
  {Doiron}}, \bibinfo {author} {\bibfnamefont {A.}~\bibnamefont
  {Litwin-Kumar}}, \bibinfo {author} {\bibfnamefont {R.}~\bibnamefont
  {Rosenbaum}}, \bibinfo {author} {\bibfnamefont {G.~K.}\ \bibnamefont
  {Ocker}}, \ and\ \bibinfo {author} {\bibfnamefont {K.}~\bibnamefont
  {Josi{\'{c}}}},\ }\href {\doibase 10.1038/nn.4242} {\bibfield  {journal}
  {\bibinfo  {journal} {Nature Neuroscience}\ }\textbf {\bibinfo {volume}
  {19}},\ \bibinfo {pages} {383} (\bibinfo {year} {2016})}\BibitemShut
  {NoStop}%
\bibitem [{\citenamefont {Ecker}\ \emph {et~al.}(2010)\citenamefont {Ecker},
  \citenamefont {Berens}, \citenamefont {Keliris}, \citenamefont {Bethge},
  \citenamefont {Logothetis},\ and\ \citenamefont {Tolias}}]{Ecker2010}%
  \BibitemOpen
  \bibfield  {author} {\bibinfo {author} {\bibfnamefont {A.~S.}\ \bibnamefont
  {Ecker}}, \bibinfo {author} {\bibfnamefont {P.}~\bibnamefont {Berens}},
  \bibinfo {author} {\bibfnamefont {G.~a.}\ \bibnamefont {Keliris}}, \bibinfo
  {author} {\bibfnamefont {M.}~\bibnamefont {Bethge}}, \bibinfo {author}
  {\bibfnamefont {N.~K.}\ \bibnamefont {Logothetis}}, \ and\ \bibinfo {author}
  {\bibfnamefont {A.~S.}\ \bibnamefont {Tolias}},\ }\href {\doibase
  10.1126/science.1179867} {\bibfield  {journal} {\bibinfo  {journal} {Science
  (New York, N.Y.)}\ }\textbf {\bibinfo {volume} {327}},\ \bibinfo {pages}
  {584} (\bibinfo {year} {2010})}\BibitemShut {NoStop}%
\bibitem [{\citenamefont {Renart}\ \emph {et~al.}(2010)\citenamefont {Renart},
  \citenamefont {Rocha}, \citenamefont {Bartho}, \citenamefont {Hollender},
  \citenamefont {Parga}, \citenamefont {Reyes},\ and\ \citenamefont
  {Harris}}]{Renart2010}%
  \BibitemOpen
  \bibfield  {author} {\bibinfo {author} {\bibfnamefont {A.}~\bibnamefont
  {Renart}}, \bibinfo {author} {\bibfnamefont {J.~D.}\ \bibnamefont {Rocha}},
  \bibinfo {author} {\bibfnamefont {P.}~\bibnamefont {Bartho}}, \bibinfo
  {author} {\bibfnamefont {L.}~\bibnamefont {Hollender}}, \bibinfo {author}
  {\bibfnamefont {N.}~\bibnamefont {Parga}}, \bibinfo {author} {\bibfnamefont
  {A.}~\bibnamefont {Reyes}}, \ and\ \bibinfo {author} {\bibfnamefont {K.~D.}\
  \bibnamefont {Harris}},\ }\href {\doibase 10.1126/science.1179850.The} {\
  \textbf {\bibinfo {volume} {327}},\ \bibinfo {pages} {587} (\bibinfo {year}
  {2010})}\BibitemShut {NoStop}%
\bibitem [{\citenamefont {Pernice}\ \emph {et~al.}(2011)\citenamefont
  {Pernice}, \citenamefont {Staude}, \citenamefont {Cardanobile},\ and\
  \citenamefont {Rotter}}]{Pernice2011}%
  \BibitemOpen
  \bibfield  {author} {\bibinfo {author} {\bibfnamefont {V.}~\bibnamefont
  {Pernice}}, \bibinfo {author} {\bibfnamefont {B.}~\bibnamefont {Staude}},
  \bibinfo {author} {\bibfnamefont {S.}~\bibnamefont {Cardanobile}}, \ and\
  \bibinfo {author} {\bibfnamefont {S.}~\bibnamefont {Rotter}},\ }\href
  {\doibase 10.1371/journal.pcbi.1002059} {\bibfield  {journal} {\bibinfo
  {journal} {PLoS Computational Biology}\ }\textbf {\bibinfo {volume} {7}}
  (\bibinfo {year} {2011}),\ 10.1371/journal.pcbi.1002059}\BibitemShut
  {NoStop}%
\bibitem [{\citenamefont {Helias}\ \emph {et~al.}(2014)\citenamefont {Helias},
  \citenamefont {Tetzlaff},\ and\ \citenamefont {Diesmann}}]{Helias2014}%
  \BibitemOpen
  \bibfield  {author} {\bibinfo {author} {\bibfnamefont {M.}~\bibnamefont
  {Helias}}, \bibinfo {author} {\bibfnamefont {T.}~\bibnamefont {Tetzlaff}}, \
  and\ \bibinfo {author} {\bibfnamefont {M.}~\bibnamefont {Diesmann}},\ }\href
  {\doibase 10.1371/journal.pcbi.1003428} {\bibfield  {journal} {\bibinfo
  {journal} {PLoS Computational Biology}\ }\textbf {\bibinfo {volume} {10}}
  (\bibinfo {year} {2014}),\ 10.1371/journal.pcbi.1003428},\ \Eprint
  {http://arxiv.org/abs/1304.2149} {arXiv:1304.2149} \BibitemShut {NoStop}%
\bibitem [{\citenamefont {Brunel}(2000)}]{Brunel2000}%
  \BibitemOpen
  \bibfield  {author} {\bibinfo {author} {\bibfnamefont {N.}~\bibnamefont
  {Brunel}},\ }\href {\doibase 10.1016/S0925-2312(00)00179-X} {\bibfield
  {journal} {\bibinfo  {journal} {Neurocomputing}\ }\textbf {\bibinfo {volume}
  {32-33}},\ \bibinfo {pages} {307} (\bibinfo {year} {2000})}\BibitemShut
  {NoStop}%
\bibitem [{\citenamefont {Pernice}\ \emph {et~al.}(2012)\citenamefont
  {Pernice}, \citenamefont {Staude}, \citenamefont {Cardanobile},\ and\
  \citenamefont {Rotter}}]{Pernice2012}%
  \BibitemOpen
  \bibfield  {author} {\bibinfo {author} {\bibfnamefont {V.}~\bibnamefont
  {Pernice}}, \bibinfo {author} {\bibfnamefont {B.}~\bibnamefont {Staude}},
  \bibinfo {author} {\bibfnamefont {S.}~\bibnamefont {Cardanobile}}, \ and\
  \bibinfo {author} {\bibfnamefont {S.}~\bibnamefont {Rotter}},\ }\href
  {\doibase 10.1103/PhysRevE.85.031916} {\bibfield  {journal} {\bibinfo
  {journal} {Physical Review E - Statistical, Nonlinear, and Soft Matter
  Physics}\ }\textbf {\bibinfo {volume} {85}},\ \bibinfo {pages} {1} (\bibinfo
  {year} {2012})},\ \Eprint {http://arxiv.org/abs/arXiv:1201.0288v2}
  {arXiv:arXiv:1201.0288v2} \BibitemShut {NoStop}%
\bibitem [{\citenamefont {Trousdale}\ \emph {et~al.}(2012)\citenamefont
  {Trousdale}, \citenamefont {Hu}, \citenamefont {Shea-Brown},\ and\
  \citenamefont {Josi\'{c}}}]{Trousdale2012}%
  \BibitemOpen
  \bibfield  {author} {\bibinfo {author} {\bibfnamefont {J.}~\bibnamefont
  {Trousdale}}, \bibinfo {author} {\bibfnamefont {Y.}~\bibnamefont {Hu}},
  \bibinfo {author} {\bibfnamefont {E.}~\bibnamefont {Shea-Brown}}, \ and\
  \bibinfo {author} {\bibfnamefont {K.}~\bibnamefont {Josi\'{c}}},\ }\href
  {\doibase 10.1371/journal.pcbi.1002408} {\bibfield  {journal} {\bibinfo
  {journal} {PLoS Computational Biology}\ }\textbf {\bibinfo {volume} {8}}
  (\bibinfo {year} {2012}),\ 10.1371/journal.pcbi.1002408},\ \Eprint
  {http://arxiv.org/abs/1110.4914} {arXiv:1110.4914} \BibitemShut {NoStop}%
\bibitem [{\citenamefont {Brunel}\ and\ \citenamefont
  {Hakim}(1999)}]{Brunel1999}%
  \BibitemOpen
  \bibfield  {author} {\bibinfo {author} {\bibfnamefont {N.}~\bibnamefont
  {Brunel}}\ and\ \bibinfo {author} {\bibfnamefont {V.}~\bibnamefont {Hakim}},\
  }\href {\doibase 10.1162/089976699300016179} {\bibfield  {journal} {\bibinfo
  {journal} {Neural Computation}\ }\textbf {\bibinfo {volume} {11}},\ \bibinfo
  {pages} {1621} (\bibinfo {year} {1999})},\ \Eprint
  {http://arxiv.org/abs/9904278} {arXiv:9904278 [cond-mat]} \BibitemShut
  {NoStop}%
\bibitem [{\citenamefont {Lindner}\ and\ \citenamefont
  {Schimansky-Geier}(2001)}]{Lindner2001}%
  \BibitemOpen
  \bibfield  {author} {\bibinfo {author} {\bibfnamefont {B.}~\bibnamefont
  {Lindner}}\ and\ \bibinfo {author} {\bibfnamefont {L.}~\bibnamefont
  {Schimansky-Geier}},\ }\href {\doibase 10.1103/PhysRevLett.86.2934}
  {\bibfield  {journal} {\bibinfo  {journal} {Physical Review Letters}\
  }\textbf {\bibinfo {volume} {86}},\ \bibinfo {pages} {2934} (\bibinfo {year}
  {2001})}\BibitemShut {NoStop}%
\bibitem [{\citenamefont {Ostojic}(2014)}]{Ostojic2014}%
  \BibitemOpen
  \bibfield  {author} {\bibinfo {author} {\bibfnamefont {S.}~\bibnamefont
  {Ostojic}},\ }\href {\doibase 10.1038/nn.3658} {\bibfield  {journal}
  {\bibinfo  {journal} {Nature neuroscience}\ }\textbf {\bibinfo {volume}
  {17}},\ \bibinfo {pages} {594} (\bibinfo {year} {2014})}\BibitemShut
  {NoStop}%
\bibitem [{\citenamefont {Jovanovi\ifmmode~\acute{c}\else \'{c}\fi{}}\ \emph
  {et~al.}(2015)\citenamefont {Jovanovi\ifmmode~\acute{c}\else \'{c}\fi{}},
  \citenamefont {Hertz},\ and\ \citenamefont {Rotter}}]{Jovanovic2015}%
  \BibitemOpen
  \bibfield  {author} {\bibinfo {author} {\bibfnamefont {S.}~\bibnamefont
  {Jovanovi\ifmmode~\acute{c}\else \'{c}\fi{}}}, \bibinfo {author}
  {\bibfnamefont {J.}~\bibnamefont {Hertz}}, \ and\ \bibinfo {author}
  {\bibfnamefont {S.}~\bibnamefont {Rotter}},\ }\href {\doibase
  10.1103/PhysRevE.91.042802} {\bibfield  {journal} {\bibinfo  {journal} {Phys.
  Rev. E}\ }\textbf {\bibinfo {volume} {91}},\ \bibinfo {pages} {042802}
  (\bibinfo {year} {2015})}\BibitemShut {NoStop}%
\bibitem [{\citenamefont {Deniz}\ and\ \citenamefont
  {Rotter}(2016)}]{Deniz2016}%
  \BibitemOpen
  \bibfield  {author} {\bibinfo {author} {\bibfnamefont {T.}~\bibnamefont
  {Deniz}}\ and\ \bibinfo {author} {\bibfnamefont {S.}~\bibnamefont {Rotter}},\
  }\href {http://arxiv.org/abs/1604.03619} {\ ,\ \bibinfo {pages} {1} (\bibinfo
  {year} {2016})},\ \Eprint {http://arxiv.org/abs/1604.03619}
  {arXiv:1604.03619} \BibitemShut {NoStop}%
\bibitem [{\citenamefont {Shadlen}\ and\ \citenamefont
  {Newsome}(1998)}]{Shadlen1998}%
  \BibitemOpen
  \bibfield  {author} {\bibinfo {author} {\bibfnamefont {M.~N.}\ \bibnamefont
  {Shadlen}}\ and\ \bibinfo {author} {\bibfnamefont {W.~T.}\ \bibnamefont
  {Newsome}},\ }\href {\doibase 0270-6474/98/183870-27$05.00/0} {\bibfield
  {journal} {\bibinfo  {journal} {The Journal of neuroscience : the official
  journal of the Society for Neuroscience}\ }\textbf {\bibinfo {volume} {18}},\
  \bibinfo {pages} {3870} (\bibinfo {year} {1998})}\BibitemShut {NoStop}%
\bibitem [{\citenamefont {Lyamzin}\ \emph {et~al.}(2015)\citenamefont
  {Lyamzin}, \citenamefont {Barnes}, \citenamefont {Donato}, \citenamefont
  {Garcia-Lazaro}, \citenamefont {Keck},\ and\ \citenamefont
  {Lesica}}]{Lyamzin2015}%
  \BibitemOpen
  \bibfield  {author} {\bibinfo {author} {\bibfnamefont {D.~R.}\ \bibnamefont
  {Lyamzin}}, \bibinfo {author} {\bibfnamefont {S.~J.}\ \bibnamefont {Barnes}},
  \bibinfo {author} {\bibfnamefont {R.}~\bibnamefont {Donato}}, \bibinfo
  {author} {\bibfnamefont {J.~A.}\ \bibnamefont {Garcia-Lazaro}}, \bibinfo
  {author} {\bibfnamefont {T.}~\bibnamefont {Keck}}, \ and\ \bibinfo {author}
  {\bibfnamefont {N.~A.}\ \bibnamefont {Lesica}},\ }\href {\doibase
  10.1523/JNEUROSCI.4738-14.2015} {\bibfield  {journal} {\bibinfo  {journal}
  {Journal of Neuroscience}\ }\textbf {\bibinfo {volume} {35}},\ \bibinfo
  {pages} {8065} (\bibinfo {year} {2015})}\BibitemShut {NoStop}%
\bibitem [{\citenamefont {Rosenbaum}\ \emph {et~al.}(2012)\citenamefont
  {Rosenbaum}, \citenamefont {Marpeau}, \citenamefont {Ma}, \citenamefont
  {Barua},\ and\ \citenamefont {Josi\'{c}}}]{Rosenbaum2012}%
  \BibitemOpen
  \bibfield  {author} {\bibinfo {author} {\bibfnamefont {R.}~\bibnamefont
  {Rosenbaum}}, \bibinfo {author} {\bibfnamefont {F.}~\bibnamefont {Marpeau}},
  \bibinfo {author} {\bibfnamefont {J.}~\bibnamefont {Ma}}, \bibinfo {author}
  {\bibfnamefont {A.}~\bibnamefont {Barua}}, \ and\ \bibinfo {author}
  {\bibfnamefont {K.}~\bibnamefont {Josi\'{c}}},\ }\href {\doibase
  10.1007/s00285-011-0451-3} {\bibfield  {journal} {\bibinfo  {journal}
  {Journal of Mathematical Biology}\ }\textbf {\bibinfo {volume} {65}},\
  \bibinfo {pages} {1} (\bibinfo {year} {2012})},\ \Eprint
  {http://arxiv.org/abs/1011.0669} {arXiv:1011.0669} \BibitemShut {NoStop}%
\bibitem [{\citenamefont {Helias}\ \emph {et~al.}(2010)\citenamefont {Helias},
  \citenamefont {Deger}, \citenamefont {Diesmann},\ and\ \citenamefont
  {Rotter}}]{Helias2010}%
  \BibitemOpen
  \bibfield  {author} {\bibinfo {author} {\bibfnamefont {M.}~\bibnamefont
  {Helias}}, \bibinfo {author} {\bibfnamefont {M.}~\bibnamefont {Deger}},
  \bibinfo {author} {\bibfnamefont {M.}~\bibnamefont {Diesmann}}, \ and\
  \bibinfo {author} {\bibfnamefont {S.}~\bibnamefont {Rotter}},\ }\href
  {\doibase 10.3389/neuro.10.029.2009} {\bibfield  {journal} {\bibinfo
  {journal} {Frontiers in computational neuroscience}\ }\textbf {\bibinfo
  {volume} {3}},\ \bibinfo {pages} {29} (\bibinfo {year} {2010})}\BibitemShut
  {NoStop}%
\bibitem [{\citenamefont {Abouzeid}\ and\ \citenamefont
  {Ermentrout}(2011)}]{Abouzeid2011}%
  \BibitemOpen
  \bibfield  {author} {\bibinfo {author} {\bibfnamefont {A.}~\bibnamefont
  {Abouzeid}}\ and\ \bibinfo {author} {\bibfnamefont {B.}~\bibnamefont
  {Ermentrout}},\ }\href {\doibase 10.1103/PhysRevE.84.061914} {\bibfield
  {journal} {\bibinfo  {journal} {Physical Review E - Statistical, Nonlinear,
  and Soft Matter Physics}\ }\textbf {\bibinfo {volume} {84}} (\bibinfo {year}
  {2011}),\ 10.1103/PhysRevE.84.061914},\ \Eprint
  {http://arxiv.org/abs/arXiv:1101.1919v1} {arXiv:arXiv:1101.1919v1}
  \BibitemShut {NoStop}%
\bibitem [{\citenamefont {Tchumatchenko}\ \emph {et~al.}(2010)\citenamefont
  {Tchumatchenko}, \citenamefont {Malyshev}, \citenamefont {Geisel},
  \citenamefont {Volgushev},\ and\ \citenamefont {Wolf}}]{Tchumatchenko2010a}%
  \BibitemOpen
  \bibfield  {author} {\bibinfo {author} {\bibfnamefont {T.}~\bibnamefont
  {Tchumatchenko}}, \bibinfo {author} {\bibfnamefont {A.}~\bibnamefont
  {Malyshev}}, \bibinfo {author} {\bibfnamefont {T.}~\bibnamefont {Geisel}},
  \bibinfo {author} {\bibfnamefont {M.}~\bibnamefont {Volgushev}}, \ and\
  \bibinfo {author} {\bibfnamefont {F.}~\bibnamefont {Wolf}},\ }\href {\doibase
  10.1103/PhysRevLett.104.058102} {\bibfield  {journal} {\bibinfo  {journal}
  {Physical Review Letters}\ }\textbf {\bibinfo {volume} {104}},\ \bibinfo
  {pages} {2} (\bibinfo {year} {2010})},\ \Eprint
  {http://arxiv.org/abs/0810.2901} {arXiv:0810.2901} \BibitemShut {NoStop}%
\bibitem [{\citenamefont {Vilela}\ and\ \citenamefont
  {Lindner}(2009)}]{Vilela2009}%
  \BibitemOpen
  \bibfield  {author} {\bibinfo {author} {\bibfnamefont {R.~D.}\ \bibnamefont
  {Vilela}}\ and\ \bibinfo {author} {\bibfnamefont {B.}~\bibnamefont
  {Lindner}},\ }\href {\doibase 10.1103/PhysRevE.80.031909} {\bibfield
  {journal} {\bibinfo  {journal} {Physical Review E - Statistical, Nonlinear,
  and Soft Matter Physics}\ }\textbf {\bibinfo {volume} {80}},\ \bibinfo
  {pages} {1} (\bibinfo {year} {2009})},\ \Eprint
  {http://arxiv.org/abs/0912.2336} {arXiv:0912.2336} \BibitemShut {NoStop}%
\bibitem [{\citenamefont {Schwalger}\ \emph {et~al.}(2015)\citenamefont
  {Schwalger}, \citenamefont {Droste},\ and\ \citenamefont
  {Lindner}}]{Schwalger2015}%
  \BibitemOpen
  \bibfield  {author} {\bibinfo {author} {\bibfnamefont {T.}~\bibnamefont
  {Schwalger}}, \bibinfo {author} {\bibfnamefont {F.}~\bibnamefont {Droste}}, \
  and\ \bibinfo {author} {\bibfnamefont {B.}~\bibnamefont {Lindner}},\ }\href
  {\doibase 10.1007/s10827-015-0560-x} {\bibfield  {journal} {\bibinfo
  {journal} {Journal of Computational Neuroscience}\ }\textbf {\bibinfo
  {volume} {39}},\ \bibinfo {pages} {29} (\bibinfo {year} {2015})}\BibitemShut
  {NoStop}%
\bibitem [{\citenamefont {Voronenko}\ \emph {et~al.}(2015)\citenamefont
  {Voronenko}, \citenamefont {{Stannat W}},\ and\ \citenamefont {{Linder
  B}}}]{Voronenko2015}%
  \BibitemOpen
  \bibfield  {author} {\bibinfo {author} {\bibfnamefont {S.~O.}\ \bibnamefont
  {Voronenko}}, \bibinfo {author} {\bibnamefont {{Stannat W}}}, \ and\ \bibinfo
  {author} {\bibnamefont {{Linder B}}},\ }\href {\doibase
  10.1186/2190-8567-5-1} {\bibfield  {journal} {\bibinfo  {journal} {The
  Journal of Mathematical Neuroscience}\ }\textbf {\bibinfo {volume} {5}},\
  \bibinfo {pages} {1} (\bibinfo {year} {2015})}\BibitemShut {NoStop}%
\bibitem [{\citenamefont {Schultze-Kraft}\ \emph {et~al.}(2013)\citenamefont
  {Schultze-Kraft}, \citenamefont {Diesmann}, \citenamefont {Gr{\"{u}}n},\ and\
  \citenamefont {Helias}}]{Schultze-Kraft2013}%
  \BibitemOpen
  \bibfield  {author} {\bibinfo {author} {\bibfnamefont {M.}~\bibnamefont
  {Schultze-Kraft}}, \bibinfo {author} {\bibfnamefont {M.}~\bibnamefont
  {Diesmann}}, \bibinfo {author} {\bibfnamefont {S.}~\bibnamefont
  {Gr{\"{u}}n}}, \ and\ \bibinfo {author} {\bibfnamefont {M.}~\bibnamefont
  {Helias}},\ }\href {\doibase 10.1371/journal.pcbi.1002904} {\bibfield
  {journal} {\bibinfo  {journal} {PLoS Computational Biology}\ }\textbf
  {\bibinfo {volume} {9}},\ \bibinfo {pages} {e1002904} (\bibinfo {year}
  {2013})},\ \Eprint {http://arxiv.org/abs/1207.7228} {arXiv:1207.7228}
  \BibitemShut {NoStop}%
\bibitem [{\citenamefont {Kuhn}\ \emph {et~al.}(2003)\citenamefont {Kuhn},
  \citenamefont {Aertsen},\ and\ \citenamefont {Rotter}}]{Kuhn2003}%
  \BibitemOpen
  \bibfield  {author} {\bibinfo {author} {\bibfnamefont {A.}~\bibnamefont
  {Kuhn}}, \bibinfo {author} {\bibfnamefont {A.}~\bibnamefont {Aertsen}}, \
  and\ \bibinfo {author} {\bibfnamefont {S.}~\bibnamefont {Rotter}},\ }\href
  {\doibase 10.1162/089976603321043702} {\bibfield  {journal} {\bibinfo
  {journal} {Neural Computation}\ }\textbf {\bibinfo {volume} {15}},\ \bibinfo
  {pages} {67} (\bibinfo {year} {2003})}\BibitemShut {NoStop}%
\bibitem [{\citenamefont {Richardson}\ and\ \citenamefont
  {Swarbrick}(2010)}]{Richardson2010}%
  \BibitemOpen
  \bibfield  {author} {\bibinfo {author} {\bibfnamefont {M.~J.~E.}\
  \bibnamefont {Richardson}}\ and\ \bibinfo {author} {\bibfnamefont
  {R.}~\bibnamefont {Swarbrick}},\ }\href {\doibase
  10.1103/PhysRevLett.105.178102} {\bibfield  {journal} {\bibinfo  {journal}
  {Phys. Rev. Lett.}\ }\textbf {\bibinfo {volume} {105}},\ \bibinfo {pages}
  {178102} (\bibinfo {year} {2010})}\BibitemShut {NoStop}%
\bibitem [{\citenamefont {Abramowitz}(1974)}]{Abramowitz:1974:HMF:1098650}%
  \BibitemOpen
  \bibfield  {author} {\bibinfo {author} {\bibfnamefont {M.}~\bibnamefont
  {Abramowitz}},\ }\href@noop {} {\emph {\bibinfo {title} {Handbook of
  Mathematical Functions, With Formulas, Graphs, and Mathematical Tables,}}}\
  (\bibinfo  {publisher} {Dover Publications, Incorporated},\ \bibinfo {year}
  {1974})\BibitemShut {NoStop}%
\bibitem [{\citenamefont {Gewaltig}\ and\ \citenamefont
  {Diesmann}(2007)}]{Gewaltig2007}%
  \BibitemOpen
  \bibfield  {author} {\bibinfo {author} {\bibfnamefont {M.-O.}\ \bibnamefont
  {Gewaltig}}\ and\ \bibinfo {author} {\bibfnamefont {M.}~\bibnamefont
  {Diesmann}},\ }\href@noop {} {\bibfield  {journal} {\bibinfo  {journal}
  {Scholarpedia}\ }\textbf {\bibinfo {volume} {2}},\ \bibinfo {pages} {1430}
  (\bibinfo {year} {2007})}\BibitemShut {NoStop}%
\bibitem [{\citenamefont {Gerstner}\ and\ \citenamefont
  {Kistler}(2002)}]{GerstnerKistler2002}%
  \BibitemOpen
  \bibfield  {author} {\bibinfo {author} {\bibfnamefont {W.}~\bibnamefont
  {Gerstner}}\ and\ \bibinfo {author} {\bibfnamefont {W.~M.}\ \bibnamefont
  {Kistler}},\ }\href {http://opac.inria.fr/record=b1099891} {\emph {\bibinfo
  {title} {Spiking neuron models : single neurons, populations, plasticity}}}\
  (\bibinfo  {publisher} {Cambridge University Press},\ \bibinfo {address}
  {Cambridge, New York, Melbourne},\ \bibinfo {year} {2002})\ \bibinfo {note}
  {autre tirage : 2006, 2008 (4e)}\BibitemShut {NoStop}%
\bibitem [{\citenamefont {Kuhn}\ \emph {et~al.}(2004)\citenamefont {Kuhn},
  \citenamefont {Aertsen},\ and\ \citenamefont {Rotter}}]{Kuhn2004a}%
  \BibitemOpen
  \bibfield  {author} {\bibinfo {author} {\bibfnamefont {A.}~\bibnamefont
  {Kuhn}}, \bibinfo {author} {\bibfnamefont {A.}~\bibnamefont {Aertsen}}, \
  and\ \bibinfo {author} {\bibfnamefont {S.}~\bibnamefont {Rotter}},\ }\href
  {\doibase 10.1523/JNEUROSCI.3349-03.2004} {\bibfield  {journal} {\bibinfo
  {journal} {Integration The Vlsi Journal}\ }\textbf {\bibinfo {volume} {24}},\
  \bibinfo {pages} {2345} (\bibinfo {year} {2004})}\BibitemShut {NoStop}%
\bibitem [{\citenamefont {de~la Rocha}\ \emph {et~al.}(2007)\citenamefont
  {de~la Rocha}, \citenamefont {Doiron}, \citenamefont {Shea-Brown},
  \citenamefont {Josi\'{c}},\ and\ \citenamefont {Reyes}}]{DelaRocha2007}%
  \BibitemOpen
  \bibfield  {author} {\bibinfo {author} {\bibfnamefont {J.}~\bibnamefont
  {de~la Rocha}}, \bibinfo {author} {\bibfnamefont {B.}~\bibnamefont {Doiron}},
  \bibinfo {author} {\bibfnamefont {E.}~\bibnamefont {Shea-Brown}}, \bibinfo
  {author} {\bibfnamefont {K.}~\bibnamefont {Josi\'{c}}}, \ and\ \bibinfo
  {author} {\bibfnamefont {A.}~\bibnamefont {Reyes}},\ }\href {\doibase
  10.1038/nature06028} {\bibfield  {journal} {\bibinfo  {journal} {Nature}\
  }\textbf {\bibinfo {volume} {448}},\ \bibinfo {pages} {802} (\bibinfo {year}
  {2007})}\BibitemShut {NoStop}%
\bibitem [{\citenamefont {Padmanabhan}\ and\ \citenamefont
  {Urban}(2010)}]{Padmanabhan2010}%
  \BibitemOpen
  \bibfield  {author} {\bibinfo {author} {\bibfnamefont {K.}~\bibnamefont
  {Padmanabhan}}\ and\ \bibinfo {author} {\bibfnamefont {N.~N.}\ \bibnamefont
  {Urban}},\ }\href@noop {} {\bibfield  {journal} {\bibinfo  {journal} {Nature
  Neuroscience}\ }\textbf {\bibinfo {volume} {13}},\ \bibinfo {pages} {1276}
  (\bibinfo {year} {2010})}\BibitemShut {NoStop}%
\bibitem [{\citenamefont {Yim}\ \emph {et~al.}(2012)\citenamefont {Yim},
  \citenamefont {Wolfart}, \citenamefont {Aertsen},\ and\ \citenamefont
  {Rotter}}]{Yim2012}%
  \BibitemOpen
  \bibfield  {author} {\bibinfo {author} {\bibfnamefont {M.~Y.}\ \bibnamefont
  {Yim}}, \bibinfo {author} {\bibfnamefont {J.}~\bibnamefont {Wolfart}},
  \bibinfo {author} {\bibfnamefont {A.}~\bibnamefont {Aertsen}}, \ and\
  \bibinfo {author} {\bibfnamefont {S.}~\bibnamefont {Rotter}},\ }\href
  {\doibase 10.3389/conf.fncom.2012.55.00030} {\bibfield  {journal} {\bibinfo
  {journal} {Frontiers in Computational Neuroscience}\ }\textbf {\bibinfo
  {volume} {6}} (\bibinfo {year} {2012}),\
  10.3389/conf.fncom.2012.55.00030}\BibitemShut {NoStop}%
\bibitem [{\citenamefont {Yim}\ \emph {et~al.}(2013)\citenamefont {Yim},
  \citenamefont {Aertsen},\ and\ \citenamefont {Rotter}}]{Yim2013}%
  \BibitemOpen
  \bibfield  {author} {\bibinfo {author} {\bibfnamefont {M.~Y.}\ \bibnamefont
  {Yim}}, \bibinfo {author} {\bibfnamefont {A.}~\bibnamefont {Aertsen}}, \ and\
  \bibinfo {author} {\bibfnamefont {S.}~\bibnamefont {Rotter}},\ }\href
  {\doibase 10.1103/PhysRevE.87.032710} {\bibfield  {journal} {\bibinfo
  {journal} {Physical Review E - Statistical, Nonlinear, and Soft Matter
  Physics}\ }\textbf {\bibinfo {volume} {87}},\ \bibinfo {pages} {1} (\bibinfo
  {year} {2013})},\ \Eprint {http://arxiv.org/abs/1208.5350} {arXiv:1208.5350}
  \BibitemShut {NoStop}%
\bibitem [{\citenamefont {Morrison}\ \emph {et~al.}(2008)\citenamefont
  {Morrison}, \citenamefont {Diesmann},\ and\ \citenamefont
  {Gerstner}}]{Morrison2008}%
  \BibitemOpen
  \bibfield  {author} {\bibinfo {author} {\bibfnamefont {A.}~\bibnamefont
  {Morrison}}, \bibinfo {author} {\bibfnamefont {M.}~\bibnamefont {Diesmann}},
  \ and\ \bibinfo {author} {\bibfnamefont {W.}~\bibnamefont {Gerstner}},\
  }\href {\doibase 10.1007/s00422-008-0233-1} {\bibfield  {journal} {\bibinfo
  {journal} {Biological Cybernetics}\ }\textbf {\bibinfo {volume} {98}},\
  \bibinfo {pages} {459} (\bibinfo {year} {2008})}\BibitemShut {NoStop}%
\bibitem [{\citenamefont {Babadi}\ and\ \citenamefont
  {Abbott}(2013)}]{Babadi2013}%
  \BibitemOpen
  \bibfield  {author} {\bibinfo {author} {\bibfnamefont {B.}~\bibnamefont
  {Babadi}}\ and\ \bibinfo {author} {\bibfnamefont {L.~F.}\ \bibnamefont
  {Abbott}},\ }\href {\doibase 10.1371/journal.pcbi.1002906} {\bibfield
  {journal} {\bibinfo  {journal} {PLoS Computational Biology}\ }\textbf
  {\bibinfo {volume} {9}},\ \bibinfo {pages} {e1002906} (\bibinfo {year}
  {2013})}\BibitemShut {NoStop}%
\bibitem [{\citenamefont {Ostojic}\ \emph {et~al.}(2009)\citenamefont
  {Ostojic}, \citenamefont {Brunel},\ and\ \citenamefont
  {Hakim}}]{Ostojic2009}%
  \BibitemOpen
  \bibfield  {author} {\bibinfo {author} {\bibfnamefont {S.}~\bibnamefont
  {Ostojic}}, \bibinfo {author} {\bibfnamefont {N.}~\bibnamefont {Brunel}}, \
  and\ \bibinfo {author} {\bibfnamefont {V.}~\bibnamefont {Hakim}},\ }\href
  {\doibase 10.1523/JNEUROSCI.1275-09.2009} {\bibfield  {journal} {\bibinfo
  {journal} {J Neurosci}\ }\textbf {\bibinfo {volume} {29}},\ \bibinfo {pages}
  {10234} (\bibinfo {year} {2009})}\BibitemShut {NoStop}%
\bibitem [{\citenamefont {Yim}\ \emph {et~al.}(2014)\citenamefont {Yim},
  \citenamefont {Kumar}, \citenamefont {Aertsen},\ and\ \citenamefont
  {Rotter}}]{Yim2014}%
  \BibitemOpen
  \bibfield  {author} {\bibinfo {author} {\bibfnamefont {M.~Y.}\ \bibnamefont
  {Yim}}, \bibinfo {author} {\bibfnamefont {A.}~\bibnamefont {Kumar}}, \bibinfo
  {author} {\bibfnamefont {A.}~\bibnamefont {Aertsen}}, \ and\ \bibinfo
  {author} {\bibfnamefont {S.}~\bibnamefont {Rotter}},\ }\href {\doibase
  10.1007/s10827-014-0502-z} {\bibfield  {journal} {\bibinfo  {journal}
  {Journal of Computational Neuroscience}\ ,\ \bibinfo {pages} {293}} (\bibinfo
  {year} {2014})}\BibitemShut {NoStop}%
\bibitem [{\citenamefont {Shea-Brown}\ \emph {et~al.}(2008)\citenamefont
  {Shea-Brown}, \citenamefont {Josi{\'{c}}}, \citenamefont {de~la Rocha},\ and\
  \citenamefont {Doiron}}]{Shea-Brown2008}%
  \BibitemOpen
  \bibfield  {author} {\bibinfo {author} {\bibfnamefont {E.}~\bibnamefont
  {Shea-Brown}}, \bibinfo {author} {\bibfnamefont {K.}~\bibnamefont
  {Josi{\'{c}}}}, \bibinfo {author} {\bibfnamefont {J.}~\bibnamefont {de~la
  Rocha}}, \ and\ \bibinfo {author} {\bibfnamefont {B.}~\bibnamefont
  {Doiron}},\ }\href {\doibase 10.1103/PhysRevLett.100.108102} {\bibfield
  {journal} {\bibinfo  {journal} {Physical Review Letters}\ }\textbf {\bibinfo
  {volume} {100}},\ \bibinfo {pages} {108102} (\bibinfo {year}
  {2008})}\BibitemShut {NoStop}%
\bibitem [{\citenamefont {Staude}\ \emph {et~al.}(2010)\citenamefont {Staude},
  \citenamefont {Rotter},\ and\ \citenamefont {Gr{\"u}n}}]{Staude2010}%
  \BibitemOpen
  \bibfield  {author} {\bibinfo {author} {\bibfnamefont {B.}~\bibnamefont
  {Staude}}, \bibinfo {author} {\bibfnamefont {S.}~\bibnamefont {Rotter}}, \
  and\ \bibinfo {author} {\bibfnamefont {S.}~\bibnamefont {Gr{\"u}n}},\ }in\
  \href@noop {} {\emph {\bibinfo {booktitle} {Analysis of Parallel Spike
  Trains}}},\ \bibinfo {editor} {edited by\ \bibinfo {editor} {\bibfnamefont
  {S.}~\bibnamefont {Rotter}}\ and\ \bibinfo {editor} {\bibfnamefont
  {S.}~\bibnamefont {Gr{\"u}n}}}\ (\bibinfo  {publisher} {Springer},\ \bibinfo
  {year} {2010})\ Chap.~\bibinfo {chapter} {12}, pp.\ \bibinfo {pages}
  {253--283}\BibitemShut {NoStop}%
\bibitem [{\citenamefont {Ames}(2014)}]{Ames2014}%
  \BibitemOpen
  \bibfield  {author} {\bibinfo {author} {\bibfnamefont {W.~F.}\ \bibnamefont
  {Ames}},\ }\href@noop {} {\emph {\bibinfo {title} {Numerical methods for
  partial differential equations}}}\ (\bibinfo  {publisher} {Academic press},\
  \bibinfo {year} {2014})\BibitemShut {NoStop}%
\end{thebibliography}%

\end{document}